\documentclass[preprint,showpacs,eqsecnum,floatfix]{revtex4-1}
\usepackage{amsmath}
\usepackage{amssymb}
\usepackage{amsfonts}
\def\beq{\begin{equation}}
\def\eeq{\end{equation}}
\def\beqa{\begin{eqnarray}}
\def\eeqa{\end{eqnarray}}

\def\btau{\mbox{\boldmath$\tau$}}

\usepackage[dvips]{graphicx,color}

\begin{document}

\title{A momentum-space Argonne V18 interaction}

\author{S. Veerasamy}
\affiliation{
Department of Physics and Astronomy, The University of Iowa, Iowa City, IA
52242}

\author{W. N. Polyzou}
\affiliation{
Department of Physics and Astronomy, The University of Iowa, Iowa City, IA
52242}

\vspace{10mm}
\date{\today}

\begin{abstract}

  This paper gives a momentum-space representation of the Argonne V18
  potential as an expansion in products of spin-isospin operators with
  scalar coefficient functions of the momentum transfer.  Two
  representations of the scalar coefficient functions for the strong
  part of the interaction are given.  One is as an expansion in an
  orthonormal basis of rational functions and the other as an
  expansion in Chebyshev polynomials on different intervals.  Both
  provide practical and efficient representations for computing the
  momentum-space potential that do not require integration or
  interpolation.  Programs based on both expansions are available as
  supplementary material.  Analytic expressions are given for the
  scalar coefficient functions of the Fourier transform of the
  electromagnetic part of the Argonne V18.  A simple method for
  computing the partial-wave projections of these interactions from
  the operator expressions is also given.
 
\end{abstract}

\vspace{10mm}

\maketitle





\section{Introduction}

The Argonne V18 potential \cite{v18} is one of a number of
nucleon-nucleon interactions \cite{reid}\cite{v18}\cite{cdbonn} that
provide a quantitative description of experimental two-body
observables below the pion-production threshold.  It is distinguished
from the other realistic interactions because it is expressed as an
operator expansion with local configuration-space coefficient
functions.  This representation has advantages when used in
variational Monte Carlo calculations.  On the other
hand, there are a number of calculations that require a realistic
interaction that are more naturally performed in momentum space.
These include some Faddeev calculations, relativistic few-body
calculations, and calculations involving electromagnetic probes.  In
the momentum representation the variable conjugate to the relative
coordinate is the momentum transfer.  In calculations, both momenta
appear, which requires either an interpolation or a separate Fourier
transform for each pair of momenta.  Fourier transforms of the V18
potential have been used in some applications \cite{huber}.  The
purpose of this paper is to provide useful, tested and
reproducible analytic approximations of the Fourier transform of the
Argonne V18 potential for use in momentum-space calculations.  The
analytic forms allow for a direct calculation of the momentum-space
interaction for any pair of initial and final momenta.  In keeping
with the traditional Argonne form, the momentum-space potential is
given as a linear combination of products of spin-isospin operators
with scalar functions of the momentum transfer.  The resulting
momentum-space potential has 24 terms.  The additional six operators
appear because the Fourier transform of the terms involving the
operators $\mathbf{L}^2 V_i(r) $ and $(\mathbf{L}\cdot \mathbf{S})^2
V_i(r)$ each become a sum of two different momentum-space operators
with different coefficient functions.  In this work the Fourier
transform is given for the strong part of the Argonne V18 potential,
without the electromagnetic terms.  This part of the potential must be
treated numerically.  The electromagnetic terms have analytic Fourier
transforms, which are discussed in Appendix ~ 3.  The partial-wave
projection of the momentum space potential is discussed in Appendix ~
2.  It is constructed from the operator expressions by integrating
over the angle between the initial and final momentum vectors, however
unlike the configuration-space partial-wave projection, the integrals
involve both the operator and the scalar coefficient functions.
 
The Argonne V18 potential has the
form
\beq 
V = \sum_{n=1}^{18} V_n (r) O_n 
\label{a.1}
\eeq 
where $V_n(r)$ are rotationally-invariant coefficient functions of the
relative coordinate of the nucleons and the $O_n$ are the 
eighteen spin-isospin operators given in Table 1.,

\vbox{
\begin{center}
Table 1: Argonne V18 spin-isospin operators \\ 
 in coordinate-space 
\vskip 2pt
\begin{tabular}{|l|c|}
\hline
Term & spin-isospin Operator in r-space \\
\hline
$O_1$ &$ \mathbf{I} $\\
$O_2 $&$ (\btau_1 \cdot \btau_2)$\\
$O_3 $&$(\pmb{\sigma}_1 \cdot \pmb{\sigma}_2),
\label{a.2}$\\
$O_4 $&$ (\pmb{\sigma}_1 \cdot
\pmb{\sigma}_2)(\btau_1 \cdot \btau_2)$\\
$O_5 $&$ S_{12}=3(\pmb{\sigma}_1 \cdot \hat{\mathbf{r}})( \pmb{\sigma}_2 \cdot
\hat{\mathbf{r}}) - \pmb{\sigma}_1 \cdot \pmb{\sigma}_2 $\\
$O_6 $&$ S_{12} (\btau_1 \cdot \btau_2),
\label{a.3} $\\
$O_7 $&$ (\mathbf{L}\cdot \mathbf{S}) $\\
$O_8 $&$
(\mathbf{L}\cdot \mathbf{S})(\btau_1 \cdot \btau_2)$\\
$ O_9  $&$ (\mathbf{L}\cdot \mathbf{L}) $\\
$O_{10} $&$(\mathbf{L}\cdot \mathbf{L})(\btau_1
\cdot \btau_2)$\\
$O_{11} $&$(\mathbf{L}\cdot
\mathbf{L})(\pmb{\sigma}_1 \cdot \pmb{\sigma}_2)$\\
$O_{12} $&$
(\mathbf{L}\cdot \mathbf{L})(\pmb{\sigma}_1 \cdot \pmb{\sigma}_2)
(\btau_1 \cdot \btau_2)
$\\
$O_{13} $&$ (\mathbf{L}\cdot \mathbf{S})^2 $\\
$O_{14} $&$ (\mathbf{L}\cdot \mathbf{S})^2 (\btau_1 \cdot \btau_2) $\\
$O_{15} $&$ T_{12} = ( 3 \tau_{1z}\tau_{2z} - \btau \cdot \btau)$\\
$O_{16} $&$ (\pmb{\sigma}_1 \cdot
\pmb{\sigma}_2)T_{12} $\\
$O_{17} $&$ S_{12} T_{12} $\\
$O_{18} $&$
(\tau_{1z} + \tau_{2z})$\\
\hline
\end{tabular}
\end{center}
}
In this table $T_{12}$ is the isotensor operator $T_{12}:= 3
\tau_{1z}\tau_{2z}- \pmb{\tau}_1 \cdot \pmb{\tau}_2 $.  While the
isospin operators, $\pmb{\tau}_i$, factor out of the Fourier
transforms, the operators $L^2$, $\mathbf{L}\cdot \mathbf{S}$,
$(\mathbf{L}\cdot \mathbf{S})^2$ and the tensor operator $S_{12}$
contribute to the Fourier transform.

The Fourier transform of this potential can be expressed as a linear
combination of 24 momentum-space operators with scalar coefficient
functions of the momentum transfer.  There are 24 operators because
the $\mathbf{L}\cdot \mathbf{L}$ and $(\mathbf{L}\cdot \mathbf{S})^2$
operators have two distinct contributions in momentum space.  In
appendix 1 it is shown that the potential matrix element $\langle
\mathbf{k}'\vert V \vert \mathbf{k} \rangle$, with
$\mathbf{q}:=\mathbf{k}' - \mathbf{k}$, has the following five types
of contributions:

\begin{itemize}  
\item[1.] \qquad $\mathbf{I}$

\beq
{1 \over (2 \pi)^3} \int e^{-i (\mathbf{k}'-\mathbf{k})\cdot \mathbf{r}}
V_j(r) \mathbf{I} d\mathbf{r} = \mathbf{I}
{1 \over 2 \pi^2} \int_0^\infty j_0 (qr) V_j(r) r^2 dr .
\label{a.8}
\eeq

\item[2.] \qquad $\mathbf{L} \cdot \mathbf{S}$

\beq
{1 \over (2 \pi)^3} \int e^{-i \mathbf{k}'\cdot \mathbf{r}}
V_j(r) \mathbf{L} \cdot \mathbf{S}
e^{i \mathbf{k}\cdot \mathbf{r}} d\mathbf{r} =
i  (\mathbf{k} \times \mathbf{k}')\cdot \mathbf{S} 
{1 \over 2 \pi^2 q} \int_0^\infty j_1 (qr) V_j(r) r^3 dr . 
\label{a.9}
\eeq

\item[3.] \qquad $\mathbf{L} \cdot \mathbf{L}$

\[
{1 \over (2 \pi)^3} \int e^{-i \mathbf{k}'\cdot \mathbf{r}}
V_j(r) \mathbf{L} \cdot \mathbf{L}
e^{i \mathbf{k}\cdot \mathbf{r}} d\mathbf{r} =
\]
\beq
- (\mathbf{k}'\times \mathbf{k}) \cdot 
(\mathbf{k}'\times \mathbf{k})
{1 \over 2 \pi^2 q^2}\int_0^\infty j_2 (qr) V_j(r) r^4 dr 
+
2 (\mathbf{k}'\cdot \mathbf{k}) 
{1 \over 2 \pi^2 q}\int_0^\infty j_1 (qr) V_j(r) r^3 dr .
\label{a.10}
\eeq

\item[4.] \qquad $(\mathbf{L} \cdot \mathbf{S})^2$

\[
{1 \over (2 \pi)^3} \int e^{-i \mathbf{k}'\cdot \mathbf{r}}
V_j(r) (\mathbf{L} \cdot \mathbf{S})^2
e^{i \mathbf{k}\cdot \mathbf{r}} d\mathbf{r} =
\]
\beq
- (\mathbf{S} \cdot ( \mathbf{k}\times \mathbf{k}'))^2
 {1 \over 2 \pi^2 q^2 } \int_0^\infty j_2 (qr) V_j(r) r^4 dr
+ 
( \mathbf{k}'\times \mathbf{S})\cdot
( \mathbf{k}\times \mathbf{S})  
{1 \over 2 \pi^2 q} \int_0^\infty j_1 (qr) V_j(r) r^3 dr .
\label{a.11}
\eeq

\item[5.] \qquad 
$S_{12}= 3(\hat{\mathbf{r}} \cdot \pmb{\sigma}_1)
(\hat{\mathbf{r}} \cdot \pmb{\sigma}_2) - \pmb{\sigma}_1\cdot \pmb{\sigma}_2$

\[
{1 \over (2 \pi)^3} \int e^{-i \mathbf{k}'\cdot \mathbf{r}}
V(r) \left (3 (\hat{\mathbf{r}} \cdot \pmb{\sigma}_1)
(\hat{\mathbf{r}} \cdot \pmb{\sigma}_2) 
-  \pmb{\sigma}_1\cdot \pmb{\sigma}_2 \right )
e^{i \mathbf{k}\cdot \mathbf{r}} d\mathbf{r} =
\]
\beq
-
\left (3 (\mathbf{q} \cdot \pmb{\sigma}_1)( \mathbf{q} \cdot \pmb{\sigma}_2)
- q^2 \pmb{\sigma}_1 \cdot \pmb{\sigma}_2 \right )
{1 \over 2 \pi^2 q^2} \int_0^\infty j_2 (qr) V(r) r^2 dr . 
\label{a.12}
\eeq
\end{itemize}
These expressions are used to represent the momentum-space interaction
as a sum of scalar functions of $q:=\vert \mathbf{q}\vert$ multiplied
by spin-isospin operators.  These scalar coefficient functions of the
momentum transfer that multiply the spin-isospin operators
have the form of one of the integrals listed in Table 2:
\begin{center}
Table 2: Momentum-space scalar coefficient functions\\
\vskip 2pt
\begin{tabular}{|l|l|l|}
\hline
Scalar coefficient function &dim& indices \\
\hline
$\tilde{V}_{m}(q) :=  {1 \over 2 \pi^2 }\int_0^\infty j_0 (qr) V_m(r) r^2 dr$
& MeV fm$^{3}$ &  $ m \in \{1,2,3,4,15,16,18 \}$ \\
$\tilde{V}_{m}(q) := {1 \over 2 \pi^2 q }\int_0^\infty j_1 (qr) V_m(r) r^3 dr$
& MeV fm$^{5}$ & $m \in \{7,8,9b,10b,11b,12b,13b,14b \}$\\
$\tilde{V}_{m}(q) :=  {1 \over 2 \pi^2 q^2}\int_0^\infty j_2 (qr) V_m(r) r^4 dr$
& MeV fm$^{7}$ &  $m \in \{9a,10a,11a,12a,13a,14a \}$ \\
$\tilde{V}_{m}(q) :=  {1 \over 2 \pi^2 q^2}\int_0^\infty j_2 (qr) V_m(r) r^2 dr$ & MeV fm$^{5}$ & $ m \in \{5,6,17 \}$\\
\hline
\end{tabular}
\end{center}

where $V_m(r)$ is the $m^{th}$ potential in the expansion (\ref{a.1})
and $\tilde{V}_{ma} (q)$ and $\tilde{V}_{mb} (q)$ are the two
different functions that appear in (\ref{a.10}) and (\ref{a.11}).
These functions have finite limits as $q\to 0$ in spite of the $1/q^l$
coefficients since the Bessel function $j_l(qr)$ vanishes like $q^l$
as $q\to 0$. The strong interaction contribution to the 24 scalar
coefficients listed in Table 2 are numerically computed. The
computational methods are discussed in section 3.  Programs that
compute these scalar coefficients are available as supplementary
material to the electronic version of this paper.  Quantities, like
the binding energies in the test calculations, exhibit small
sensitivities (in the sixth significant figure) to the precision of
input constants. In the supplementary programs these constants are
taken from the original V18 potential.

The electromagnetic contribution to each of these operators can be
represented in terms of known special functions.  These contributions
are important for precise low-energy calculations and can be added to
the strong interaction coefficient functions when they are needed.
The analytic expressions for the electromagnetic terms are given in Appendix ~2.

The resulting momentum-space potential has an operator expansion
of the form
\beq
\langle \mathbf{k}' \vert V \vert \mathbf{k} \rangle = 
\sum_{m\in S}  \tilde{V}_{m} (q) \tilde{O}_m 
\label{a.17}
\eeq
where $S=\{1,2,3,4,5,6,7,8,9a,9b,10a,10b,11a,11b,12a,12b,13a,13b,14a,14b,15,16,17$,$18\}$ and the 24 operators $\tilde{O}_m $ are given in Table 3.

\begin{center}
\vbox{
Table 3: Argonne V18  momentum-space \\
spin-isospin operators  \\ 
\vskip 2pt 
\begin{tabular}{|l|l|}
\hline
term & spin-isospin operator \\
\hline
$\tilde{O}_{1}$ & $ \mathbf{I} $ \\
$\tilde{O}_{2}$ & $ (\btau_1 \cdot \btau_2) $\\
$\tilde{O}_{3}$ & $ (\pmb{\sigma}_1 \cdot \pmb{\sigma}_2)$\\
$\tilde{O}_{4}$ & $(\pmb{\sigma}_1 \cdot \pmb{\sigma}_2)(\btau_1 \cdot \btau_2)$\\ 
$\tilde{O}_{5}$ & $ - \left (3 (\mathbf{q} \cdot \pmb{\sigma}_1)( \mathbf{q} \cdot \pmb{\sigma}_2)
- {q^2 } \pmb{\sigma}_1 \cdot \pmb{\sigma}_2 \right ) $\\
$
\tilde{O}_{6} $ & $
- \left (3 (\mathbf{q} \cdot \pmb{\sigma}_1)( \mathbf{q} \cdot \pmb{\sigma}_2)
- {q^2 } \pmb{\sigma}_1 \cdot \pmb{\sigma}_2 \right )
(\btau_1 \cdot \btau_2) $\\
$
\tilde{O}_{7} $ & $
i (\mathbf{k} \times \mathbf{k}')\cdot \mathbf{S} $\\  
$ \tilde{O}_{8} $ & $
i (\mathbf{k} \times \mathbf{k}') \cdot \mathbf{S} (\btau_1 \cdot \btau_2)
$\\
$
\tilde{O}_{9a} $ & $
-(\mathbf{k}'\times \mathbf{k}) \cdot 
(\mathbf{k}'\times \mathbf{k})  
$ \\
$
\tilde{O}_{9b} $ & $
2 (\mathbf{k}'\cdot \mathbf{k}) $\\ 
$
\tilde{O}_{10a} $ & $
-(\mathbf{k}'\times \mathbf{k}) \cdot 
(\mathbf{k}'\times \mathbf{k})  (\btau_1 \cdot \btau_2)
$\\
$
\tilde{O}_{10b} $ & $
2 (\mathbf{k}'\cdot \mathbf{k})(\btau_1 \cdot \btau_2) 
$\\
$
\tilde{O}_{11a}$ & $
-(\mathbf{k}'\times \mathbf{k}) \cdot 
(\mathbf{k}'\times \mathbf{k}) (\pmb{\sigma}_1 \cdot \pmb{\sigma}_2) 
$\\
$
\tilde{O}_{11b} $ & $
2(\mathbf{k}'\cdot \mathbf{k})(\pmb{\sigma}_1 \cdot \pmb{\sigma}_2)  
$\\
$
\tilde{O}_{12a} $ & $
-(\mathbf{k}'\times \mathbf{k}) \cdot 
(\mathbf{k}'\times \mathbf{k})  (\pmb{\sigma}_1 \cdot
\pmb{\sigma}_2) (\btau_1 \cdot \btau_2) 
$\\
$
\tilde{O}_{12b} $ & $
2 (\mathbf{k}'\cdot \mathbf{k})(\pmb{\sigma}_1 \cdot
\pmb{\sigma}_2) (\btau_1 \cdot \btau_2) 
$\\
$
\tilde{O}_{13a} $ & $
- (\mathbf{S} \cdot ( \mathbf{k}\times \mathbf{k}'))^2
$\\
$\tilde{O}_{13b} $ & $
( \mathbf{k}'\times \mathbf{S})\cdot
( \mathbf{k}\times \mathbf{S}) 
$\\
$
\tilde{O}_{14a} $ & $
- (\mathbf{S} \cdot ( \mathbf{k}\times \mathbf{k}'))^2
 (\btau_1 \cdot \btau_2) 
$\\
$
\tilde{O}_{14b} $ & $
( \mathbf{k}'\times \mathbf{S})\cdot
( \mathbf{k}\times \mathbf{S}) (\btau_1 \cdot \btau_2)
$\\
$\tilde{O}_{15}  $ & $
T_{12} 
$\\
$
\tilde{O}_{16} $ & $
(\pmb{\sigma}_1 \cdot \pmb{\sigma}_2) T_{12} 
$ \\
$
\tilde{O}_{17} $ & $
- \left (3 (\mathbf{q} \cdot \pmb{\sigma}_1)( \mathbf{q} \cdot \pmb{\sigma}_2)
- {q^2 } \pmb{\sigma}_1 \cdot \pmb{\sigma}_2 \right )T_{12}
$\\
$
\tilde{O}_{18} $ & $
(\tau_{1z} + \tau_{2z}). 
$\\
\hline
\end{tabular}
}
\end{center}

The Argonne V18 potential in momentum-space has the dimension $MeV
fm^{3}$. Dividing by $\hbar c$ in Mev-fermi can be used to convert the
momentum-space potential to a consistent set of units, $(fm)^{2}$.

\section{Numerical Fourier Bessel transforms}

This section summarizes an accurate numerical computation of the
integrals in Table 2.  These computations are used to test the accuracy of
the approximations discussed in the next section.

Because the configuration-space potential falls off asymptotically
like $e^{-m_{\pi} r}$, the radial integrals are evaluated with a
finite cutoff at 20 $fm$.  The Fourier-Bessel transforms are evaluated
for momentum transfers $q<100\,fm^{-1}$.  With these cutoffs the
maximum value of $x:=qr$ that can appear in the argument of the
spherical Bessel functions in the integrals in Table 2. is
$x_{max}=2000$.  To evaluate these integrals the zeros of the
spherical Bessel functions $j_0(x)$, $j_1(x)$, and $j_2(x)$ for $0\leq
x \leq 2000$ are computed for each fixed value of $q$.  For each value
of $q$ the integrals are expressed as sums of integrals between
successive zeros of the spherical Bessel function that appear in the
integral.  If $q$ is such that $qr$ is never a zero of $j_l(qr)$ for
$0<r<20 fm$ then the integral over $r$ is performed using a 100 point
Gauss-Legendre quadrature on the interval $[0,20fm]$.  If $q$ is such
that $qr$ has zeros of $j_l(qr)$ for $0<r<20fm$, then the integrals
between zeros $[qr_i , qr_{i+1}]$ are computed using 20 Gauss-Legendre
points when $r_{i+1} \leq 5\, fm$, 40 Gauss-Legendre points when
$5\,fm< r_{i+1} \leq 10\,fm$ and 80 Gauss-Legendre points when
$10\,fm< r_{i+1} \leq 20\,fm$. For further details see \cite{thesis2011}

\section{Approximations}

This section discusses two approximations of the potential
functions $\tilde{V}_{m}
(q)$ in Table 2 by expansions in known elementary functions.
The first method approximates these potential functions by linear
combinations of Chebyshev polynomials on three distinct intervals of
momenta, for momenta up to 100$fm^{-1}$.  The second approach
approximates these potential functions by a finite linear combination
of orthonormal functions of the momentum transfer that have analytic
Fourier-Bessel transforms.  The configuration-space basis functions
are associated Laguerre polynomials multiplied by decaying
exponentials.  These functions have analytic Fourier transforms that
are rational functions of the momentum transfer \cite{keister}.  In
both approaches the coefficients of the expansion function are stored.
The basis functions at any point can be generated efficiently by
recursion and the potentials can be expressed as a finite linear
combination of the basis functions.  Both methods lead to efficient
and accurate approximations to the momentum-space potential.

Figures 1 and 2 show the potential functions for the central and
tensor parts ($V_1(q)$ and $V_5(q)$) of the interaction to illustrate
the structure of typical potentials.

\subsection{Chebyshev expansions}

This section discusses the Chebyshev basis.
The functions $\tilde{V}_m(q)$ are replaced by a Chebyshev polynomial 
approximation on the interval $q \in [a,b]$ using \cite{broucke}
\beq
\tilde{V}_m(q) \approx 
c_0/2 +\sum_{n=1}^{100} c_n T_n (-{a+b\over b-a}+ {2 \over b-a}q)
\label{aa.1}
\eeq
where 
\beq 
T_{n}(x) = \cos(n \cos^{-1} (x)))
\label{aa.2}
\eeq 
are Chebyshev polynomials and the coefficients $c_n$ are
computed using a Clenshaw-Curtiss quadrature \cite{broucke} :
\beq
c_n = {2 \over N} 
[
{1 \over 2}\tilde{V}_m(b)+ 
\sum_{j=1}^{N-1} 
\tilde{V}_m({a+b\over 2} + {b-a \over 2} \cos (\pi j /N))  \cos (nj \pi /N)+
(-)^n {1 \over 2} \tilde{V}_m(a)
]
\label{aa.3}
\eeq
with $N=101$.  The functions $\tilde{V}_m (q)$ are evaluated at the
quadrature points $q_j:={a+b\over 2} + {b-a \over 2} \cos (\pi j /N)$
using the methods discussed above.  This is repeated for $q$ in each
of three intervals, $[a,b]= [0,10], [10,50], [50,100]$ and the 101
expansion coefficients associated with each of these three intervals
are stored.  The Chebyshev polynomials are computed using the
recurrence relations
\beq
T_{n+1} (x) = 2x T_n(x) - T_{n-1}(x), \qquad T_0 (x)=1, \, T_1 (x) = x .
\label{aa.4}
\eeq
For $q$ larger than 100 $fm^{-1}$ $\tilde{V}_m(q)$ is approximated by 0.

For the potentials $\tilde{V}_4(q)$, $\tilde{V}_6(q)$ 
and $\tilde{V}_{17}(q)$ it was
necessary to add additional Chebyshev expansions intervals between
zero and ten $fm^{-1}$.  For $\tilde{V}_4(q)$ 21 polynomials were used
on $[0,.2]fm^{-1}$, 31 polynomials were used on $[.2,.5]fm^{-1}$, 41
polynomials were used on $[.5,2.0]fm^{-1}$ and 71 polynomials were
used on $[2.0,10.0]fm^{-1}$.  For $\tilde{V}_6(q)$ 31
polynomials were used on $[0,.5]fm^{-1}$, 41 polynomials were used on
$[.5,2.0]fm^{-1}$ and 41 polynomials were used on $[2.0,10.0]fm^{-1}$.
Similarly
for $\tilde{V}_{17}(q)$ 31
polynomials were used on $[0,1.0]fm^{-1}$, 51 polynomials were used on
$[1.0,5.0]fm^{-1}$ and 51 polynomials were used on $[5.0,10.0]fm^{-1}$.

This method provides an accurate and efficient representation for
computing a momentum space $V18$ interaction.  One of the
supplementary programs (chebyshev-argonne.c) uses this method to
compute the 24 coefficient functions in Table 2.

\subsection{Rational basis functions} 

While the method of the previous section gives accurate results, a
more straightforward approach is to represent the potential directly
as an expansion in basis functions that have analytic Fourier
transforms.  In order to represent the potential, each of the scalar
potentials $\tilde{V}_m (q)$, is approximated by an expansion in known
basis functions.  A method to compute both the expansion coefficients
and a recursion formula to compute basis functions are given below.

The functions $V_m(r)$, $rV_m(r)$, and $r^2V_m(r)$ that appear in the
integrands of the integrals in Table 2 are expanded using an
orthonormal set of radial functions that have analytic Fourier-Bessel
transforms \cite{keister}.  These functions are associated Laguerre
polynomials multiplied by decaying exponentials in configuration
space.  Their Fourier-Bessel transforms have power-law fall of in
momentum space.  In addition, they vanish at the origin in a manner
that can be used to explicitly cancel the factors $1/q$ and $1/q^2$
that appear in the definitions of
$\tilde{V}_{m}$ in Table 2.  Both sets of basis functions can be generated
efficiently using recursion relations.  The cancellation of the
factors $1/q$ and $1/q^2$ can be directly incorporated into the
recursion that generates the momentum-space basis functions so the
final expression for the potential does not require a special
treatment for $q$ near $0$.

The radial basis functions for different values of $l$ are given
below.  The dimensionless parameter $x := \Lambda r$ is used in the
basis functions, where $\Lambda$ is a scale parameter that can be
chosen to improve efficiency.  The parameterization of the Argonne
V18 interaction uses the value $\Lambda = 7(fm)^{-1}$.  The
configuration-space basis functions are
\beq
\phi_{nl}(r) = {1 \over \sqrt{N_{nl}}}x^l L_n^{2l+2}(2x) e^{-x} 
\label{b.1}
\eeq
where 
\beq
L_n^\alpha = \sum_{m=0}^n (-)^m 
\left (
\begin{array}{c}
n+\alpha \\
n-m \\
\end{array}
\right )
{x^m \over m!}
\label{b.2}
\eeq
and the normalization coefficient is 
\beq
N_{nl} = \Lambda^{-3} ({1 \over 2})^{2l+3} {\Gamma (n+\alpha+1 ) \over n!}.
\label{b.3}
\eeq
These functions satisfy the orthogonality relations
\beq
\int_0^\infty 
\phi_{nl}(r) \phi_{ml}(r) r^2 dr = \delta_{mn} .
\label{b.4}
\eeq
They have analytic Fourier-Bessel transforms given by 
\beq
\tilde{\phi}_{nl}(q) = \sqrt{2 \over \pi} \int_0^{\infty} 
j_l(qr) \phi_{nl} (r) r^2 dr .
\label{b.5}
\eeq
For $y=q / \Lambda$ the $\tilde{\phi}_{nl}(q)$ can be expressed in terms
of Jacobi polynomials:
\beq
\tilde{\phi}_{nl}(q) =
{1 \over \sqrt{\tilde{N}_{nl}}}{y^l \over (y^2 +1)^{l+2}}
P_n^{l+{3 \over 2},l+{1 \over 2}}({y^2-1 \over y^2+1})
\label{b.6}
\eeq
with normalization coefficient
\beq
\tilde{N}_{nl} = {\Lambda^3 \over 2 (2n+2l+3)}
{\Gamma(n+l+{5 \over 2})\Gamma(n+l+{3 \over 2}) \over
n! \Gamma (n+2l+3)}
\label{b.7}
\eeq
and
\beq
P^{(\alpha,\beta)}_n (x) = 
{\Gamma (\alpha+n+1) \over n! \Gamma (\alpha +\beta +n +1)}
\sum_{m=0}^n \left ( 
\begin{array}{c}
n\\
m
\end{array}
\right ) 
{\Gamma (\alpha+\beta +n +m+1) \over 2^m \Gamma (\alpha+m +1)}
(x-1)^m .
\label{b.8}
\eeq
These functions satisfy the orthogonality relations
\beq
\int_0^\infty 
\tilde{\phi}_{nl}(q) \tilde{\phi}_{ml}(q) q^2 dq = \delta_{mn} .
\label{b.9}
\eeq

These basis functions can be generated by using the recursion
formulas for the associated Laguerre functions and Jacobi
polynomials 
\beq
(n+1) L^\alpha_{n+1} (x) = 
(2n+\alpha +1 -x)  L^\alpha_{n+} (x)
- (n+\alpha)  L^\alpha_{n-1} (x)
\label{b.12}
\eeq
and 
\[
2(n+1)(n+\alpha+\beta +1) (2n+\alpha + \beta) 
P_{n+1} ^{(\alpha,\beta)} (x) = 
\]
\[
[(2n+\alpha +\beta +1)(\alpha^2 -\beta^2)
+ x((2n+\alpha+\beta)(2n+\alpha+\beta+1)(2n+\alpha+\beta+2)]
P_{n} ^{(\alpha,\beta)} (x)
\]
\beq 
- 
2(n+\alpha)(n+\beta)(2n+\alpha+\beta+2) 
P_{n-1} ^{(\alpha,\beta)} (x) .
\label{b.13}
\eeq
These recursion relations can be modified to incorporate the 
normalization constants (\ref{b.3}) and ({\ref{b.7}) directly 
into the recursion.  The recursion for the 
normalized radial basis functions with $(x=\Lambda r)$ is given by: 
\beq
\phi_{0l} (r) = {1 \over \sqrt{(2 l+1)!}}{1 \over \sqrt{2^{2l+3}}}
\Lambda^{3/2} x^l e^{-x} 
\label{b.14}
\eeq
\beq
\phi_{1l} (r)  =  {2l+3-2x \over \sqrt{2l+3}} \phi_{0l} (r) 
\label{b.15}
\eeq
\beq
\phi_{nl} (r) = {2n+1+2l-2x \over \sqrt{n+1+2l}\sqrt{n}} \phi_{n-1,l} (r)
- 
\sqrt{{(n-1)(n+1+2l) \over n (n+2+2l)}} \phi_{n-2,l} (r) .
\label{b.16}
\eeq
Similarly, the normalized momentum-space basis functions with 
$(y=q/\Lambda)$ are generated by the recursion:
\[
\tilde{\phi}_{0l} (q) =
\]
\beq
{1 \over \sqrt{(2l+3)!}}{1 \over \sqrt{\Lambda^3}}
{1 \over \sqrt{{1 \over 2} \cdots {2l+3 \over 2}}}
{1 \over \sqrt{{1 \over 2} \cdots {2l+1 \over 2}}} y^l {1 \over (y^2 +1)^{2l+2}}
\label{b.17}
\eeq
\beq
\tilde{\phi}_{1l} (q) = 
({1 \over 2} + (l+2){y^2-1 \over y^2+1} )
\sqrt{{2l+5 \over (l+2 +{1 \over 2})(l+1 +{1 \over 2})}}
\tilde{\phi}_{0l} (q) 
\label{b.18}
\eeq 
\[
\tilde{\phi}_{nl} (q) =
\] 
\[
\sqrt{{(2n + 2l +3)n (n+2l+2)\over 
(2n+2l +1)(n+l +{3 \over 2})(n+l +{1 \over 2}) }}
\]
\[
\times {(2n+2l+1)(2l+2)(y^2+1) + (2n+2l+1)(2n+2l)(2n+2l+2)(y^2-1)  
\over 
2n(n+2l+2)(2n+2l)(y^2-1)
}\tilde{\phi}_{n-1l} (q)
\]
\[
-
\sqrt{{(2n + 2l +3)(n-1)n(n+1+2l)(n+2l+2)\over 
(2n+2l -1)(n+l +{3 \over 2})(n+l +{1 \over 2})
(n+l +{1 \over 2})(n+l -{1 \over 2})
}}
\]
\beq
\times
{(n+l+{1 \over 2}) (n+l-{1 \over 2}) (2n+2l+2) \over
(n)(n+2l+2)(2n+2l)
}\tilde{\phi}_{n-1l} (q).
\label{b.19}
\eeq

Replacing $\tilde{\phi}_{0l} (q)$ in (\ref{b.17}) by
$\hat{\phi}_{0l} (q):= \tilde{\phi}_{0l} (q)/q^l$ given by
\[
\hat{\phi}_{0l} (q) =
\]
\beq
{1 \over \sqrt{(2l+3)!}}{1 \over \sqrt{\Lambda^3}}
{1 \over \sqrt{{1 \over 2} \cdots {2l+3 \over 2}}}
{1 \over \sqrt{{1 \over 2} \cdots {2l+1 \over 2}}} \Lambda^{-l} 
{1 \over (y^2 +1)^{2l+2}}
\label{b.20}
\eeq
to start the recursion in equations (\ref{b.18})-(\ref{b.19})
generates $\hat{\phi}_{nl} (q):= \tilde{\phi}_{nl} (q)/q^l$, which are
well-behaved as $q \to 0$.  Seventy expansion coefficients are used to
construct the momentum-space potential for each value of $m$
\beq
c_{nm}=  {1 \over 2 \pi^2 }\int_0^\infty \phi_{n0} (r) V_m(r) r^2 dr
\qquad m \in \{1,2,3,4,15,16,18 \}
\label{b.21}
\eeq
\beq
c_{nm} = {1 \over 2 \pi^2  }\int_0^\infty \phi_{n1} (r) V_m(r) r^3 dr
\qquad m \in \{7,8,9b,10b,11b,12b,13b,14b \}
\label{b.22}
\eeq
\beq
c_{nm} =  {1 \over 2 \pi^2 }\int_0^\infty \phi_{n2} (r) V_m(r) r^4 dr
\qquad m \in \{9a,10a,11a,12a,13a,14a \}
\label{b.23}
\eeq
\beq
c_{nm} =  {1 \over 2 \pi^2 }\int_0^\infty \phi_{n2} (r) V_m(r) r^2 dr
\qquad m \in \{5,6,17 \}.
\label{b.24}
\eeq
The integrals are approximated using an 80 point Gauss-Legendre
quadrature between 0 and 10$fm$.  The basis functions $\phi_{nl}(r)$ are
generated using (\ref{b.14}-\ref{b.16}).  The scale parameter in the
recursion for $\phi_{nl}(r)$ is taken as $\Lambda =7fm^{-1} $.

The 70x24 expansion coefficients $c_{nm}$ are stored.  The 
momentum-space potential functions are then given by 
\beq
\tilde{V}_{m}(q) = \sum_{n=1}^{70} c_{nm} \hat{\phi}_{nl} (q) 
\label{b.25}
\eeq
where the reduced expansion functions $\hat{\phi}_{nl}(q) := 
\tilde{\phi}_{nl}(q)/q^l$ are generated recursively using 
(\ref{b.18}-\ref{b.20}).

The full momentum-space potential in operator form is given by
\beq
V = \sum_{m\in S} \tilde{V}_{m}(q) \tilde{O}_{m} 
\eeq
where $\tilde{O}_{m}$ are the 24 operators in Table 3
and $q=\sqrt{k^2 + k^{\prime 2} - 2 \mathbf{k}' \cdot \mathbf{k}}$.

One of the supplementary programs (rational-argonne.c) uses this method to
compute the 24 coefficient functions in Table 2.

\section{Tests} 

Two tests are performed on the potentials.  First, the momentum-space
coefficient functions, $\tilde{V}_m(q)$, computed using the 
accurate numerical Fourier Bessel transforms, the Chebyshev expansion
and the rational basis function expansion are compared.
For the second test both representations of potential are used to
compute the deuteron binding energy and wave functions.  These results
are compared to a direct calculation of these quantities using the
partial-wave expansion of the original configuration space potential.

The results of the first test are shown in Tables 4-7, which list
values of the Fourier-Bessel transforms of the 24 radial functions
computed using these three different methods for momentum transfers of
1,5,15 and 25 $fm^{-1}$.

These results are shown in Tables 4,5,6 and 7 for all 24 operators and
a representative range of the momentum transfers. The columns labeled
RFExp show the scalar potential functions using the rational function
expansion, the columns labeled CExp show the same quantities using the
Chebyshev expansion, while the columns labeled NFT show the results of
the direct numerical Fourier transform.  Figures 3-26 plot the
difference of the approximate Fourier transforms with an accurate
Fourier Bessel transform divided by half of the sum of these
quantities.  The solid curves are for the rational function expansion
and the dotted curved are for the Chebyshev expansion.

\pagebreak
\vbox{
\begin{center}
Table 4: Values of scalar coefficients  at 1 fm$^{-1}$ \\
\vskip 2pt
\begin{tabular}{rrrr}
\hline
n &  RFExp  & CExp & NFT\\
\hline
1    &   6.789973e-01  &    6.789977e-01 &  6.789977e-01  \\
2    &  -4.019392e-01  &   -4.019392e-01 & -4.019392e-01  \\
3    &  -1.692090e-01  &   -1.692090e-01 & -1.692090e-01  \\
4    &   2.358519e-01  &    2.356704e-01 &  2.356705e-01  \\
5    &   7.216739e-03  &    7.218217e-03 &  7.218217e-03  \\
6    &   2.857732e-01  &    2.860471e-01 &  2.860467e-01  \\
7    &  -5.511547e-01  &   -5.511547e-01 & -5.511547e-01  \\
8    &  -1.678888e-01  &   -1.678888e-01 & -1.678888e-01  \\
9    &   1.741415e-01  &    1.741415e-01 &  1.741415e-01  \\
10   &  -3.272988e-02  &   -3.272987e-02 & -3.272987e-02  \\
11   &   1.999136e-02  &    1.999136e-02 &  1.999136e-02  \\
12   &  -7.414060e-03  &   -7.414060e-03 & -7.414060e-03  \\
13   &   9.084422e-02  &    9.084424e-02 &  9.084424e-02  \\
14   &   1.245017e-01  &    1.245017e-01 &  1.245017e-01  \\
15   &   1.122388e-02  &    1.122389e-02 &  1.122389e-02  \\
16   &  -1.214926e-02  &   -1.216021e-02 & -1.216031e-02  \\
17   &   2.403290e-03  &    2.420818e-03 &  2.420818e-03  \\
18   &   6.124964e-03  &    6.124964e-03 &  6.124964e-03  \\
19   &   1.304278e-02  &    1.304274e-02 &  1.304274e-02  \\
20   &  -1.702409e-02  &   -1.702401e-02 & -1.702401e-02  \\
21   &  -7.227244e-03  &   -7.227256e-03 & -7.227256e-03  \\
22   &  -7.849686e-03  &   -7.849707e-03 & -7.849707e-03  \\
23   &   4.518193e-02  &    4.518262e-02 &  4.518262e-02  \\
24   &   3.980251e-02  &    3.980269e-02 &  3.980269e-02  \\
\hline                                              
\end{tabular}
\end{center} 
}

\vbox{                         
\begin{center}
Table 5: Value of scalar coefficients at 5 fm$^{-1}$ \\
\vskip 2pt
\begin{tabular}{rrrr}
\hline
n & RFExp & CExp & NFT \\
\hline
1  &  1.160699e+00   &   1.160699e-00 &  1.160699e-00 \\
2  & -1.360382e-02   &  -1.360382e-02 & -1.360382e-02 \\
3  & -1.148807e-01   &  -1.148807e-01 & -1.148807e-01 \\
4  & -1.065288e-01   &  -1.065203e-01 & -1.065203e-01 \\
5  &  4.489757e-03   &   4.489763e-03 &  4.489763e-03 \\
6  &  4.405849e-03   &   4.405371e-03 &  4.405370e-03 \\
7  &  -4.623736e-02  &  -4.623736e-02 & -4.623736e-02 \\
8  &  -1.871380e-02  &  -1.871380e-02 & -1.871380e-02 \\
9  &   2.471311e-02  &   2.471311e-02 &  2.471311e-02 \\
10 &   1.480758e-03  &   1.480758e-03 &  1.480758e-03 \\
11 &   6.027203e-03  &   6.027203e-03 &  6.027203e-03 \\
12 &   1.465070e-03  &   1.465070e-03 &  1.465070e-03 \\
13 &   5.222260e-03  &   5.222260e-03 &  5.222260e-03 \\
14 &   8.233502e-03  &   8.233502e-03 &  8.233502e-03 \\
15 &   4.828280e-03  &   4.828280e-03 &  4.828280e-03 \\
16 &  -4.815794e-03  &  -4.815305e-03 & -4.815305e-03 \\
17 &   1.656921e-06  &   1.627533e-06 &  1.627533e-06 \\
18 &   4.274306e-04  &   4.274306e-04 &  4.274306e-04 \\
19 &   4.273833e-03  &   4.273832e-03 &  4.273832e-03 \\
20 &   1.791462e-04  &   1.791461e-04 &  1.791461e-04 \\
21 &   9.672551e-04  &   9.672550e-04 &  9.672550e-04 \\
22 &   1.814761e-04  &   1.814761e-04 &  1.814761e-04 \\
23 &   1.620319e-03  &   1.620318e-03 &  1.620318e-03 \\
24 &   1.790086e-03  &   1.790085e-03 &  1.790085e-03 \\
\hline
\end{tabular}
\end{center} 
}
\vbox{
\begin{center}
Table 6: Value of scalar coefficients at 15 fm$^{-1}$ \\
\vskip 2pt
\begin{tabular}{rrrr}
\hline
n & RFExp & CExp & NFT\\
\hline
1    &    9.321365e-04  &   9.321031e-04 &  9.321031e-04 \\
2    &    4.123439e-05  &   4.123387e-05 &  4.123387e-05 \\
3    &   -1.924812e-05  &  -1.924669e-05 & -1.924669e-05 \\
4    &   -6.648375e-05  &  -6.643904e-05 & -6.643770e-05 \\
5    &   -9.010902e-06  &  -9.010512e-06 & -9.010512e-06 \\
6    &    1.026393e-05  &   1.026324e-05 &  1.026323e-05 \\
7    &    5.541260e-06  &   5.540856e-06 &  5.540856e-06 \\
8    &    2.632043e-06  &   2.631901e-06 &  2.631901e-06 \\
9    &   -1.962835e-06  &  -1.962585e-06 & -1.962585e-06 \\
10   &   -9.304609e-07  &  -9.304846e-07 & -9.304846e-07 \\
11   &   -6.015901e-07  &  -6.015363e-07 & -6.015363e-07 \\
12   &   -1.047669e-07  &  -1.047529e-07 & -1.047529e-07 \\
13   &   -4.725022e-06  &  -4.725152e-06 & -4.725152e-06 \\
14   &   -1.527634e-06  &  -1.527584e-06 & -1.527584e-06 \\
15   &    2.942747e-06  &   2.942623e-06 &  2.942623e-06 \\
16   &   -2.895027e-06  &  -2.892432e-06 & -2.892244e-06 \\
17   &   -2.865458e-10  &  -3.049671e-10 & -3.061292e-10 \\
18   &    9.986465e-08  &   9.985107e-08 &  9.985107e-08 \\
19   &   -2.604487e-07  &  -2.604660e-07 & -2.604660e-07 \\
20   &   -6.335039e-08  &  -6.334951e-08 & -6.334951e-08 \\
21   &   -7.055132e-08  &  -7.055521e-08 & -7.055521e-08 \\
22   &   -1.454468e-08  &  -1.454569e-08 & -1.454569e-08 \\
23   &   -3.115148e-07  &  -3.115089e-07 & -3.115089e-07 \\
24   &   -1.394089e-07  &  -1.394129e-07 & -1.394129e-07 \\
\hline
\end{tabular}
\end{center} 
}

\vbox{
\begin{center}
Table 7: Value of scalar coefficients at 25 fm$^{-1}$ \\
\vskip 2pt
\begin{tabular}{rrrr}
\hline
n & RFExp & CExp & NFT\\
\hline
1   &  -1.386301e-05  &  -1.383431e-05  & -1.383431e-05  \\
2   &  -6.108349e-08  &  -6.010007e-08  & -6.010007e-08  \\
3   &   8.598072e-07  &   8.595154e-07  &  8.595154e-07  \\
4   &   1.014189e-06  &   1.003839e-06  &  1.003915e-06  \\
5   &  -4.600082e-07  &  -4.599210e-07  & -4.599210e-07  \\
6   &   4.739733e-07  &   4.738710e-07  &  4.738711e-07  \\
7   &   2.443040e-08  &   2.442088e-08  &  2.442088e-08  \\
8   &   9.428095e-09  &   9.412965e-09  &  9.412965e-09  \\
9   &  -1.534834e-08  &  -1.533919e-08  & -1.533919e-08  \\
10  &   3.457372e-10  &   3.607579e-10  &  3.607579e-10  \\
11  &  -3.619628e-09  &  -3.613201e-09  & -3.613201e-09  \\
12  &  -1.005784e-09  &  -1.003137e-09  & -1.003137e-09  \\
13  &   4.666338e-09  &   4.709390e-09  &  4.709390e-09  \\
14  &  -3.274714e-09  &  -3.270324e-09  & -3.270324e-09  \\
15  &  -5.425469e-08  &  -5.415686e-08  & -5.415686e-08  \\
16  &   5.452722e-08  &   5.398602e-08  &  5.399643e-08  \\
17  &  -2.888773e-12  &  -4.263050e-12  & -4.213031e-12  \\
18  &  -5.852151e-09  &  -5.841440e-09  & -5.841440e-09  \\
19  &  -2.512190e-10  &  -2.555613e-10  & -2.555613e-10  \\
20  &   7.827015e-12  &   7.411785e-12  &  7.411785e-12  \\
21  &  -5.864134e-11  &  -5.980761e-11  & -5.980761e-11  \\
22  &  -1.617550e-11  &  -1.652711e-11  & -1.652711e-11  \\
23  &   8.297311e-11  &   8.271942e-11  &  8.271943e-11  \\
24  &  -5.322500e-11  &  -5.424546e-11  & -5.424546e-11  \\
\hline
\end{tabular}
\end{center} 
}

These tables show generally good agreement among the three methods of
computation.  At 1$fm^{-1}$ and 5$fm^{-1}$ the Chebyshev expansion
agrees with the direct numerical Fourier transform to between 5-7
significant figures for all 24 potentials.  There is similar agreement
at 15$fm^{-1}$ and 25$fm^{-1}$ except in potentials 17.  The
agreement between the potentials calculated using the rational
function expansion do not agree with the direct numerical Fourier
transforms as well as the Chebyshev expansion.  The accuracy depends
on the particular potential and gets worse as the momentum transfer
increases.  Thus for precision calculations the Chebyshev expansion is
preferred.

Figures 1-24 provide a more complete picture of the nature of the
errors in both approximations.  Spikes in the errors occur near points
where the potentials change sign.  Some of the errors near zero are
enhanced because the some of the plotted potential are divided by
powers of the momentum transfer.  For these terms the operators
include compensating powers of the momentum transfer that vanish near
the origin, so the contribution of the error in the full potential
near the origin is reduced.  The rational function expansions have
larger relative errors near higher and lower values of the momentum
transfer.  This is not surprising because the basis functions are not
local.  The Chebyshev expansion is uniformly good, in part because it
is a local expansion, so more intervals can be added as needed.  The
largest errors are in potential 17.  At $10 fm^{-1}$ its value is
about $-1.2 \times 10^{-8}$, which is several orders of magnitude
smaller than any of the other potentials at that momentum transfer.

\begin{center}
Table 8: deuteron $s$ and $d$ wave functions using Chebyshev expansion,\\
rational function expansion and $r$-space partial waves \\
\vskip 2pt
\begin{tabular}{r|rrr|rrr}
\hline
k$fm^{-1}$ & $u_s(k)$-CExp.  & $u_s(k)$-RFExp  & $u_s(k)$-pw & $u_d(k)$-CExp, & $u_d(k)$-RFExp & $u_d(k)$-pw\\
\hline
0.0 &   1.2695e+01 &  1.2695e+01 &  1.2693e+01  &  0.00000e+00 &  0.00000e+00 &  0.00000e+00\\ 
0.5 &   1.9609e+00 &  1.9609e+00 &  1.9609e+00  & -2.19827e-01 & -2.19811e-01 & -2.19826e-01\\
1.0 &   3.7684e-01 &  3.7685e-01 &  3.7684e-01  & -1.72164e-01 & -1.72159e-01 & -1.72164e-01\\
1.5 &   8.2472e-02 &  8.2472e-02 &  8.2471e-02  & -1.12429e-01 & -1.12429e-01 & -1.12429e-01\\
2.0 &   6.0809e-03 &  6.0808e-03 &  6.0806e-03  & -7.10857e-02 & -7.10863e-02 & -7.10857e-02\\
2.5 &  -1.3615e-02 & -1.3616e-02 & -1.3615e-02  & -4.45428e-02 & -4.45432e-02 & -4.45428e-02\\	
3.0 &  -1.6153e-02 & -1.6153e-02 & -1.6153e-02  & -2.76853e-02 & -2.76854e-01 & -2.76853e-02\\	 
3.5 &  -1.3648e-02 & -1.3648e-02 & -1.3648e-02  & -1.69880e-02 & -1.69881e-02 & -1.69881e-02\\ 
4.0 &  -1.0153e-02 & -1.0153e-02 & -1.0153e-02  & -1.02233e-02 & -1.02234e-02 & -1.02234e-02\\
4.5 &  -6.9954e-03 & -6.9954e-03 & -6.9953e-03  & -5.98472e-03 & -5.98475e-03 & -5.98472e-03\\
5.0 &  -4.5270e-03 & -4.5270e-03 & -4.5270e-03  & -3.37040e-03 & -3.37043e-03 & -3.37041e-03\\
\hline                                                                             
\end{tabular}
\end{center} 
                                                                           
As a second test the deuteron binding energy and wave functions are
computed using the two different momentum-space potentials are
compared to each other and to the direct Fourier transform of wave
functions computed using a configuration-space partial-wave
calculation.

The deuteron binding energy and the $s$ and $d$ wave
functions are computed using the operator form of the Fourier transformed
potential, by direct integration of the vector variables.  The method
of solution, which is discussed in \cite{charlotte}, uses the
expansion (\ref{a.17}) directly without using partial waves.
Calculations are performed for both the Chebyshev and
rational-function representations of the momentum space potentials.

These calculations are compared to a configuration-space partial-wave
calculation.  In that calculation, labeled $pw$ in table 9, the wave
functions are represented by an expansion in 70 configuration-space
basis functions using the configuration-space basis functions (\ref{b.1}).
Matrix elements of the partial wave projection of the Hamiltonian,
with the configuration space Argonne V18 potential, are computed in
this basis and the eigenvalue problem is solved numerically.  The
Fourier transform is computed by analytically Fourier transforming the
basis functions. The solution of the eigenvalue problem gives an
independent evaluation of both the binding energy and wave functions
constructed directly from the configuration space potential.
  
The deuteron binding energy obtained using the Chebyshev
representation of the Fourier transform gives a deuteron binding energy of
$E_d=-2.242233$ MeV. The rational function representation gives a d
deuteron binding energy of $E_d=-2.242193$ MeV compared with
$E_d=-2.242211$ MeV using the configuration space partial-wave
calculation.  The binding energies based on all three calculation
agree to within 22 eV.  The computation used in the configuration-space 
partial-wave calculation is a Galerkin calculation and thus
gives a variational bound on the binding energy. 

These eigenvalues differ from the experimental deuteron binding
energy.  This is because the momentum-space potentials used in these
computations do not include electromagnetic corrections that appear in
the Argonne V18 codes.  The electromagnetic corrections contribute an
additional $+17.6$ keV \cite{wiringa} to the binding energy which is
consistent with the experimental binding energy of $-2.2246 $ MeV.

The $s$ and $d$ wave functions for all three calculations are compared
in Table 10.  The wave functions differ in the fifth or sixth
significant figure, while binding energies of all three calculations
differ in the sixth significant figure.

The electromagnetic contributions to the Argonne V18 potential are
important for low-energy high-precision calculations.  The Fourier
transform of the electromagnetic contributions of the Argonne V18
potential can be computed analytically, and can be added to the strong
interaction contribution discussed in this paper when necessary.  The
analytic Fourier transform of the electromagnetic contribution is
discussed in appendix 3.

While this paper gives the momentum-space version of the operator
expansion of the Argonne V18 potential, it is often useful to have
partial-wave contributions of the potential.  These can be computed
from the operator matrix elements using a one-dimensional integration
over the cosine of the angle between the initial and final 
momenta.  A simple method to compute the partial-wave
projections from the operator expansion is given in the appendix 2.

The programs to compute the potentials $\tilde{V}_m(q)$ using both
methods are available as supplementary material to the electronic
version of this article.

This work supported by the U.S. Department of Energy, contract \#
DE-FG02-86ER40286 and National Science Foundation grants
NSF-PHY-1005578 and NSF-PHY-1005501.  The authors would like to acknowledge
useful comments from Prof. C. Elster in preparing this manuscript.

\vfill\eject

\section{Appendix 1} In this appendix we compute the 
Fourier transform of the parts of the potential containing the five 
types of operators, 
$\mathbf{I}$,
$\mathbf{L}\cdot \mathbf{S}$,
$\mathbf{L}\cdot \mathbf{L}$,$(\mathbf{L}\cdot \mathbf{S})^2$, and $S_{12}$
that appear in equations(\ref{a.9}-\ref{a.12}).

\begin{center}
{\bf \noindent $\mathbf{L} \cdot \mathbf{S}$:} 
\end{center}
Let $\mathbf{q}:= \mathbf{k}'-\mathbf{k}$. 

The Fourier transform is
\[
{1 \over (2 \pi)^3} \int e^{-i \mathbf{k}'\cdot \mathbf{r}}
V_j(r) \mathbf{L} \cdot \mathbf{S}
e^{i \mathbf{k}\cdot \mathbf{r}} d\mathbf{r} =
{1 \over (2 \pi)^3} \int e^{-i \mathbf{k}'\cdot \mathbf{r}}
V_j(r) \mathbf{S} \cdot (\mathbf{r}\times \mathbf{p}) 
e^{i \mathbf{k}\cdot \mathbf{r}} d\mathbf{r} =
\]
\[
{1 \over (2 \pi)^3} \int e^{-i \mathbf{k}'\cdot \mathbf{r}}
V_j(r) \mathbf{S} \cdot (\mathbf{r}\times \mathbf{k}) 
e^{i \mathbf{k}\cdot \mathbf{r}} d\mathbf{r} =
{1 \over (2 \pi)^3} \int e^{-i \mathbf{q}\cdot \mathbf{r}}
V_j(r) \mathbf{S} \cdot (\mathbf{r}\times \mathbf{k}) 
d\mathbf{r} =
\]
\beq
{4 \pi \over (2 \pi)^3} \int
\sum_{l=0}^\infty \sum_{m=-l}^l
(-i)^l j_l (qr) Y_{lm} (\hat{\mathbf{q}})Y_{lm}^* (\hat{\mathbf{r}}) 
V_j(r) \mathbf{S} \cdot (\mathbf{r}\times \mathbf{k}) 
d\mathbf{r}.
\label{ap.1}
\eeq
Since $\mathbf{r}$ can be expanded as a linear combination of
spherical harmonics,
$Y_{1m}(\hat{\mathbf{r}})$, the only terms that survive are the $l=1$
terms.  The integral over angles and the spherical harmonics 
replace $\hat{\mathbf{r}}$ by $\hat{\mathbf{q}}$, giving
\[
 -{4\pi i \over (2 \pi)^3}
\int_{0}^\infty
j_1 (qr)  
V_j(r) \mathbf{S} \cdot (\mathbf{q}\times \mathbf{k}) 
r^3 dr =
\]
\beq
i \mathbf{S} \cdot (\mathbf{k}\times \mathbf{k}')
\times [
{ 1 \over 2 \pi^2 q}
\int_{0}^\infty
j_1 (qr)  
V_j(r)  
r^3 dr ] .
\label{ap.2}
\eeq
Thus the Fourier transform has the form
\beq
{1 \over (2 \pi)^3} \int e^{-i \mathbf{k}'\cdot \mathbf{r}}
V_j(r) \mathbf{L} \cdot \mathbf{S}
e^{i \mathbf{k}\cdot \mathbf{r}} d\mathbf{r} =
i \mathbf{S} \cdot (\mathbf{k}\times \mathbf{k}') 
I_1(q)
\label{ap.3}
\eeq
where 
\beq
I_1 (q) := { 1 \over 2 \pi^2 q}
\int_{0}^\infty j_1 (qr) V_j(r)  
r^3 dr . 
\label{ap.4}
\eeq

The following relations are used to compute Fourier transforms of 
the remaining three operators:
\beq
\pmb{\nabla}_q f(q) = f'(q) {\mathbf{q} \over q}
\eeq
\beq
\nabla_q^2 f(q)  =
f''(q) + {2 \over q} f'(q) 
\eeq
\[
(\mathbf{a} \cdot \pmb{\nabla}_q)
(\mathbf{b} \cdot \pmb{\nabla}_q) f(q) = 
\]
\beq
{\mathbf{a} \cdot \mathbf{q} \over q}{\mathbf{b} \cdot \mathbf{q} \over q}(
f''(q) - {f'(q) \over q} ) + {\mathbf{a} \cdot \mathbf{b} \over q} f'(q) . 
\eeq

\begin{center}
{\bf  $\mathbf{L} \cdot \mathbf{L}$:}
\end{center}
The Fourier transform of this operator is 
\[
{1 \over (2 \pi)^3} \int e^{-i \mathbf{k}'\cdot \mathbf{r}}
V_j(r) (\mathbf{r}\times \mathbf{p}) \cdot (\mathbf{r}\times \mathbf{p})
e^{i \mathbf{k}\cdot \mathbf{r}} d\mathbf{r} =
{1 \over (2 \pi)^3} \int e^{-i \mathbf{k}'\cdot \mathbf{r}}
V_j(r) (\mathbf{r}\times \mathbf{k}') \cdot (\mathbf{r}\times \mathbf{k})
e^{i \mathbf{k}\cdot \mathbf{r}} d\mathbf{r} =
\]
\beq
(i\pmb{\nabla}_{q}\times \mathbf{k}') 
\cdot (i\pmb{\nabla}_q\times \mathbf{k})
{1 \over (2 \pi)^3} \int V_j(r) 
e^{-i \mathbf{q}\cdot \mathbf{r}} d\mathbf{r} =
-(\pmb{\nabla}_{q}\times \mathbf{k}') 
\cdot (\pmb{\nabla}_{q}\times \mathbf{k})
{4\pi \over (2 \pi)^3} \int_0^\infty V_j(r) 
j_0 (qr) r^2 dr .
\eeq
To compute the derivatives use
\[
(\pmb{\nabla}_{q}\times \mathbf{k}') 
\cdot (\pmb{\nabla}_{q}\times \mathbf{k}) =
(\mathbf{k} \cdot \mathbf{k}')(\pmb{\nabla}_{q}\cdot \pmb{\nabla}_{q})
-
(\mathbf{k} \cdot \pmb{\nabla}_{q} )
(\mathbf{k} \cdot \pmb{\nabla}_{q} )
\]
in the above to get 
\beq
-(\pmb{\nabla}_{q}\times \mathbf{k}') 
\cdot (\pmb{\nabla}_{q}\times \mathbf{k})
{4\pi \over (2 \pi)^3} \int_0^\infty V_j(r) 
j_0 (qr) r^2 dr = 
- \left ((\mathbf{k}'\cdot \mathbf{k})\nabla^2_q
- (\mathbf{k}'\cdot \pmb{\nabla}_{q})(
\mathbf{k}\cdot \pmb{\nabla}_{q}) \right )I_0 (q)
\eeq
where
\beq
I_0(q)= {4\pi \over (2 \pi)^3} \int_0^\infty V_j(r) 
j_0 (qr) r^2 dr =
{1 \over 2 \pi^2} \int_0^\infty V_j(r) j_0 (qr) r^2 dr .
\eeq
Evaluating this gives
\[
- \left ((\mathbf{k}'\cdot \mathbf{k})\nabla^2_q
- (\mathbf{k}'\cdot \pmb{\nabla}_{q})(
\mathbf{k}\cdot \pmb{\nabla}_{q}) \right )I_0 (q)
=
\]
\beq
-(\mathbf{k}'\cdot \mathbf{k})(I_0''(q) + {1 \over q}I_0'(q)) +
{(\mathbf{k}'\cdot \mathbf{q})(\mathbf{k}\cdot \mathbf{q})\over q^2}
(I_0''(q)- {1 \over q}I_0'(q)).
\eeq
To eliminate the derivatives use 
\[
I_0''(q)-  {1 \over q}I_0'(q) =
\]
\[
{1 \over 2 \pi^2} \int_0^{\infty} V(r) (j_0'' (qr) - j_0' (qr) {1\over qr} ) r^4 dr =
{1 \over 2 \pi^2} \int_0^{\infty} V(r) j_2 (qr) r^4 dr = I_2(q) 
\]
and 
\[
I_0''(q)+  {1 \over q}I_0'(q) = 
{1 \over 2 \pi^2} \int_0^{\infty} V(r) (j_0'' (qr) - j_0' (qr) {1\over qr}
+ 2 j_0' (qr) {1\over qr} 
 ) r^4 dr =
\]
\[
{1 \over 2 \pi^2} \int_0^{\infty} V(r) j_2'' (qr) r^4 dr
- {1 \over 2 \pi^2} {2\over q}  \int_0^{\infty} V(r) j_1 (qr) r^3 dr 
=
I_2(q) - {2\over q}I_1(q).
\]
This gives
\beq
{1 \over (2 \pi)^3} \int e^{-i \mathbf{k}'\cdot \mathbf{r}}
V_j(r) (\mathbf{r}\times \mathbf{p}) \cdot (\mathbf{r}\times \mathbf{p})
e^{i \mathbf{k}\cdot \mathbf{r}} d\mathbf{r} =
-(\mathbf{k}'\cdot \mathbf{k})(I_2(q) - {2\over q}I_1(q)) +
{(\mathbf{k}'\cdot \mathbf{q})( \mathbf{k}\cdot \mathbf{q} )\over q^2}
I_2(q)
\eeq
which can be reexpressed in terms of cross products:
\beq
{1 \over (2 \pi)^3} \int e^{-i \mathbf{k}'\cdot \mathbf{r}}
V_j(r) (\mathbf{r}\times \mathbf{p}) \cdot (\mathbf{r}\times \mathbf{p})
e^{i \mathbf{k}\cdot \mathbf{r}} d\mathbf{r} =
- I_2(q) {(\mathbf{k}'\times \mathbf{k}) \cdot 
(\mathbf{k}'\times \mathbf{k}) \over q^2} + {2\over q}
(\mathbf{k}'\cdot \mathbf{k}) I_1(q) . 
\eeq

\begin{center}
{\bf $(\mathbf{L} \cdot \mathbf{S})^2$: }
\end{center}
The Fourier transform is
\[
{1\over (2\pi)^3} 
\int e^{-i(\mathbf{k}'-\mathbf{k})\cdot \mathbf{r}} V_j (r) 
(\mathbf{L} \cdot \mathbf{S})^2 d \mathbf{r} =
{1\over (2\pi)^3} 
\int e^{-i(\mathbf{k}'-\mathbf{k})\cdot \mathbf{r}} V_j (r) 
(\mathbf{S} \cdot (\mathbf{r} \times \mathbf{p}))^2 d \mathbf{r} =
\]
\[
-{4\pi\over (2\pi)^3} 
(( \mathbf{k}'\times \mathbf{S})
\cdot \pmb{\nabla}_q)
(( \mathbf{k}\times \mathbf{S})
\cdot \pmb{\nabla}_q)
\int j_0(qr) V_j (r) r^2 dr =
\]
\[
- (( \mathbf{k}'\times \mathbf{S})
\cdot \pmb{\nabla}_q)
(( \mathbf{k}\times \mathbf{S})
\cdot \pmb{\nabla}_q) I_0 (q) =
\]
\[
- (( \mathbf{k}'\times \mathbf{S})\cdot \mathbf{q}
(( \mathbf{k}\times \mathbf{S})\cdot \mathbf{q} ))
{1 \over q^2} (I_0'' (q) - {1 \over q} I'_0 (q))
- 
( \mathbf{k}'\times \mathbf{S})\cdot
( \mathbf{k}\times \mathbf{S}){1 \over q} I'_0 (q) =
\]

\beq
- (( \mathbf{k}'\times \mathbf{S})\cdot \mathbf{q}
(( \mathbf{k}\times \mathbf{S})\cdot \mathbf{q} ))=
{1 \over q^2} I_2 (q)
+ 
( \mathbf{k}'\times \mathbf{S})\cdot
( \mathbf{k}\times \mathbf{S}){1 \over q}  I_1 (q)
\eeq
which gives
\beq
- ((\mathbf{S} \cdot ( \mathbf{k}\times \mathbf{k}'))^2 
{1 \over q^2} I_2 (q)
+ 
( \mathbf{k}'\times \mathbf{S})\cdot
( \mathbf{k}\times \mathbf{S}){1 \over q}  I_1 (q)
\eeq
or
\beq
{1\over (2\pi)^2} 
\int e^{-i(\mathbf{k}'-\mathbf{k})} V_j (r) 
(\mathbf{L} \cdot \mathbf{S})^2 d \mathbf{r} =
- ((\mathbf{S} \cdot ( \mathbf{k}\times \mathbf{k}'))^2
{1 \over q^2} I_2 (q)
+ 
( \mathbf{k}'\times \mathbf{S})\cdot
( \mathbf{k}\times \mathbf{S}){1 \over q}  I_1 (q) .
\eeq

\begin{center}
{\bf ${1 \over 3} S_{12}= 
\left (\hat{\mathbf{r}} \cdot \pmb{\sigma}_1)
(\hat{\mathbf{r}} \cdot \pmb{\sigma}_2) 
- {1 \over 3} \pmb{\sigma}_1\cdot \pmb{\sigma}_2 \right )$:}
\end{center}
The Fourier transform is 
\[
{1 \over (2 \pi)^3} \int e^{-i \mathbf{k}'\cdot \mathbf{r}}
V(r) \left ( (\hat{\mathbf{r}} \cdot \pmb{\sigma}_1)
(\hat{\mathbf{r}} \cdot \pmb{\sigma}_2) 
- {1 \over 3} \pmb{\sigma}_1\cdot \pmb{\sigma}_2 \right )
e^{i \mathbf{k}\cdot \mathbf{r}} d\mathbf{r} =
\]
\[
-\left ( (\pmb{\nabla}_q \cdot \pmb{\sigma}_1)
(\pmb{\nabla}_q \cdot \pmb{\sigma}_2) 
- {1 \over 3} \pmb{\sigma}_1\cdot \pmb{\sigma}_2 \pmb{\nabla}_q^2 \right )
{4\pi \over (2 \pi)^3} \int
V(r){r^2 \over r^2} j_0(qr)  dr =
\]
\[
-{\pmb{\sigma}_1 \cdot \mathbf{q} \over q}
{\pmb{\sigma}_2 \cdot \mathbf{q} \over q}(
(I_{0-}''(q) - {I_{0-}'(q) \over q} ) - 
{\mathbf{s}_1 \cdot \mathbf{s}_2 \over q} I_0'(q) =  
\]
\beq
+{1 \over 3} \pmb{\sigma}_1\cdot \pmb{\sigma}_2 
(I''_{0-}(q) + {2 \over q} I_{0-}'(q))
\eeq
Thus,
\beq
{1 \over (2 \pi)^3} \int e^{-i \mathbf{k}'\cdot \mathbf{r}}
V(r) \left (3 (\hat{\mathbf{r}} \cdot \pmb{\sigma}_1)
(\hat{\mathbf{r}} \cdot \pmb{\sigma}_2) 
-  \pmb{\sigma}_1\cdot \pmb{\sigma}_2 \right )
e^{i \mathbf{k}\cdot \mathbf{r}} d\mathbf{r} =
-\left ( 
3{\pmb{\sigma}_1 \cdot \mathbf{q} \over q}
{\pmb{\sigma}_2 \cdot \mathbf{q} \over q} 
- 
\pmb{\sigma}_1 \cdot \pmb{\sigma}_2 
\right ) 
I_{2-}(q)
\eeq
where 
\beq
I_{0-}(q) = {4\pi \over (2 \pi)^3} \int
V(r) j_2(qr) r^2 dr . 
\eeq

\section{Appendix 2} 

In this appendix we compute the partial-wave projection of potentials
from the vector representation of the momentum-space potential.  Using
rotational invariance the partial-wave potentials can be expressed
using a one-dimensional integral over the cosine of the angle between
the initial and final momentum vectors.  The method below is similar
to the method first proposed in \cite{ce11}.

Rotational invariance of the potential implies
\[
\langle j, \mu, k, l, s \vert V \vert j', \mu', k', l', s' \rangle
\]
\[
=\langle j, \mu, k, l, s \vert U^{\dagger}(R) V U(R) \vert j', \mu', k', l', s' \rangle
\]
\[
=\sum_{\nu \nu'}D^{j*}_{\nu \mu}(R)  D^{j}_{\nu' \mu'}(R)  
\langle j, \nu, k, l, s \vert V \vert j', \nu', k', l', s' \rangle
\]
\[
=\int dR \sum_{\nu \nu'}
D^{j*}_{\nu \nu'}(R)  D^{j}_{\nu' \mu'}(R)  
\langle j, \nu, k, l, s \vert V \vert j', \nu', k', l', s' \rangle
\]
\[
= \delta_{jj'} \sum_{\mu' \nu'} \delta_{\mu \mu'}{\delta_{\nu \nu'} \over 2j+1} 
\langle j, \nu, k, l, s \vert V \vert j', \nu', k', l', s' \rangle
\]
\beq
= \delta_{\mu \mu'} \delta_{jj'} 
\langle  k, l, s \Vert  V^j \Vert k', l', s' \rangle
\label{ap2.1}
\eeq
where we have integrated over the $SU(2)$ Haar measure with normalization 
$\int dR =1 $ 
and defined the partial-wave potentials by
\beq
\langle  k, l, s \Vert  V^j \Vert k', l', s' \rangle 
:={1 \over 2j+1} \sum_{\mu=-j}^j \sum_{m_lm_l'}\sum_{m_s m_s'} 
\langle j, \mu, k, l, s \vert V \vert j, \mu, k', l', s' \rangle .
\label{ap2.2}
\eeq
This kernel is rotationally invariant.  Formally the partial-wave
potential is 
\[
\langle  k, l, s \Vert  V^j \Vert k', l', s' \rangle
\]
\[
=\int d \hat{\mathbf{k}} d\hat{\mathbf{k}}' {1 \over 2j+1} \sum_{\mu=-j}^j
\langle {1\over 2}, \mu_1, {1 \over 2}, \mu_2 \vert s, m_s \rangle 
\langle l, m_l, s, m_s \vert j, \mu \rangle 
Y_{lm_l}^* (\hat{\mathbf{k}}) 
\]
\[
\times
\langle \mathbf{k}, \mu_1 , \mu_2 \Vert V \Vert \mathbf{k}', \mu_1', 
\mu_2' \rangle
\]
\beq
\times
\langle {1\over 2}, \mu_1', {1 \over 2}, \mu_2' \vert s', m_s' \rangle 
\langle l', m_l', s', m_s' \vert j, \mu \rangle 
Y_{l'm_l'}(\hat{\mathbf{k}}').
\label{ap2.3}
\eeq
For any fixed rotation, $R$, rotational invariance of $V$ gives
\[
\langle \mathbf{k}, \mu_1 , \mu_2 \Vert V \Vert \mathbf{k}', \mu_1', 
\mu_2' \rangle 
\]
\[
=\langle \mathbf{k}, \mu_1 , \mu_2 \Vert U^{\dagger}(R)  V U(R)  
\Vert \mathbf{k}', \mu_1', 
\mu_2' \rangle
\]
\beq
=\sum_{\mu_1''\mu_2''}\sum_{\mu_1'''\mu_2'''}
D^{*1/2}_{\mu_1'' \mu_1}(R) 
D^{*1/2}_{\mu_2'' \mu_2}(R) 
\langle R\mathbf{k}, \mu_1'' , \mu_2'' \Vert V \Vert R\mathbf{k}', \mu_1''', 
\mu_2''' \rangle
D^{1/2}_{\mu_1''' \mu_1'}(R) 
D^{1/2}_{\mu_2''' \mu_2'}(R). 
\label{ap2.4}
\eeq
Using this expression in eq.~(\ref{ap2.3})  gives 
\[
\langle  k, l, s \Vert  V^j \Vert k', l', s' \rangle
\]
\[ 
=\int d \hat{\mathbf{k}} d\hat{\mathbf{k}}' {1 \over 2j+1}
\sum_{\mu=-j}^j \langle {1\over 2}, \mu_1, {1 \over 2}, \mu_2 \vert s,
m_s \rangle \langle l, m_l, s, m_s \vert j \mu \rangle \langle
Y^*_{lm_l}(\hat{\mathbf{k}}) \times 
\]
\[ 
D^{*1/2}_{\mu_1'' \mu_1}(R) D^{*1/2}_{\mu_2'' \mu_2}(R) 
\langle R\mathbf{k}, \mu_1'' , \mu_2'' \Vert V \Vert R\mathbf{k}', \mu_1''',
\mu_2''' \rangle D^{1/2}_{\mu_1''' \mu_1'}(R) D^{1/2}_{\mu_2'''
\mu_2'}(R) \times
\]
\beq 
\langle {1\over 2}, \mu_1', {1 \over 2}, \mu_2' \vert s',
m_s' \rangle \langle l', m_l', s', m_s' \vert j \mu \rangle \langle
Y_{l'm_l'}(\hat{\mathbf{k}}') .
\label{ap2.5}
\eeq
Next we eliminate three of the integrals in the potential matrix.
For any fixed $\mathbf{k}'$ there is an $R$ such that $R (k') \hat{\mathbf{k}}'
= \mathbf{\hat{z}}$.  Obviously $R^{-1} (k')=R_z(\phi') R_y (\theta')
R_z(\xi)$ where $(\theta',\phi')$ are the polar angles of $\mathbf{k}'$
and $\xi$ is arbitrary has this property where
\beq
R_y (\theta) = 
\left (
\begin{array}{ccc}
\cos (\theta ) & 0 & \sin (\theta) \\
0 & 1 &  \\
-\sin (\theta ) & 0 & \cos (\theta) \\
\end{array}
\right )
\label{ap2.6}
\eeq
\beq
R_z (\phi) = 
\left (
\begin{array}{ccc}
\cos (\phi ) &  -\sin (\phi)  & 0\\
\sin (\phi ) & \cos (\phi) & 0 \\
0 & 0 & 1 \\
\end{array}
\right ).
\label{ap2.7}
\eeq
In this case
\beq
R \mathbf{k}'= k' \hat{\mathbf{z}}
\label{ap2.8}
\eeq
and 
\beq
R \mathbf{k} = R^{-1}_z(\xi) R^{-1}_y (\theta') R^{-1}_z(\phi') \mathbf{k} .
\label{ap2.9}
\eeq
For fixed $\mathbf{k}'$ define $\mathbf{k}''$ by 
\beq
\mathbf{k}'' = R^{-1}_y (\theta') R^{-1}_z(\phi') \mathbf{k} .
\label{ap2.10}
\eeq
Given these fixed (primed) angles we change the unprimed integration 
variables 
$\mathbf{k}\to \mathbf{k}''$.  We also have 
\beq
R \mathbf{k} = R^{-1}_z(\xi) \mathbf{k}''
\label{ap2.11}
\eeq
We are also free to choose the angle $\xi$ in $R^{-1}_z(\xi)$.  We
choose it so it transforms $\mathbf{k}''$ to the $x-z$ plane.  This is
achieved by letting $\xi$ be the azimuthal angle of $\mathbf{k'}''$
\beq
R_z^{-1}  (\phi'') = 
\left (
\begin{array}{ccc}
\cos (\phi'' ) &  \sin (\phi'')  & 0\\
-\sin (\phi'' ) & \cos (\phi'') & 0 \\
0 & 0 & 1 \\
\end{array}
\right )
\label{ap2.11b}
\eeq
\beq
R_z^{-1}  (\phi'') \mathbf{k}'' = 
\left (
\begin{array}{ccc}
\cos (\phi'' ) &  \sin (\phi'')  & 0\\
-\sin (\phi'' ) & \cos (\phi'') & 0 \\
0 & 0 & 1 \\
\end{array}
\right )
\left (
\begin{array}{c}
k''\sin (\theta'') \cos (\phi'' ) \\
k''\sin (\theta'' ) \sin (\phi'')  \\
k''\cos (\theta'')  \\
\end{array}
\right )=
\left (
\begin{array}{c}
k''\sin (\theta'') \\
0  \\
k'' \cos (\theta'')  \\
\end{array}
\right ).
\label{ap2.12}
\eeq
With these substitutions the partial wave integral becomes
\[
\langle  k, l, s \Vert  V^j \Vert k', l', s' \rangle
\]
\[ 
=\int d \hat{\mathbf{k}}'' d\hat{\mathbf{k}}' {1 \over 2j+1}
\sum_{\mu=-j}^j 
\langle {1\over 2}, \mu_1, {1 \over 2}, \mu_2 \vert s, m_s \rangle 
\langle l, m_l, s, m_s \vert j, \mu \rangle 
Y^*_{lm} (R^{-1} R_z^{-1}(\phi'') \hat{\mathbf{k}}'') 
\]
\[ 
\times 
D^{*1/2}_{\mu_1'' \mu_1}(R) D^{*1/2}_{\mu_2'' \mu_2}(R) 
\langle 
k''(\hat{\mathbf{x}}\sin (\theta'')+\hat{\mathbf{z}}\cos(\theta'')) , 
\mu_1'',\mu_2'' \Vert V \Vert k'\hat{\mathbf{z}}, \mu_1''',
\mu_2''' \rangle 
D^{1/2}_{\mu_1''' \mu_1'}(R) D^{1/2}_{\mu_2''' \mu_2'}(R)
\]
\[ 
\times 
\langle {1\over 2}, \mu_1', {1 \over 2}, \mu_2' \vert s', m_s' \rangle 
\langle l', m_l', s', m_s' \vert j, \mu \rangle \langle
Y_{l'm_l'} (R^{-1} \hat{\mathbf{z}}' )
\]
\[ 
=\int d \hat{\mathbf{k}}'' d\hat{\mathbf{k}}' {1 \over 2j+1}
\sum_{\mu=-j}^j 
\langle {1\over 2}, \mu_1, {1 \over 2}, \mu_2 \vert s, m_s \rangle 
\langle l, m_l, s, m_s \vert j, \mu \rangle 
Y_{lm}^* (\hat{\mathbf{x}}\sin (\theta'')+\hat{\mathbf{z}}\cos(\theta''))  
D^{l*}_{m_l'' m_l} (R) 
\]
\[ 
\times 
D^{*1/2}_{\mu_1'' \mu_1}(R) D^{*1/2}_{\mu_2'' \mu_2}(R) 
\langle 
k''(\hat{\mathbf{x}}\sin (\theta'')+\hat{\mathbf{z}}\cos(\theta'') , 
\mu_1'',\mu_2'' \Vert V \Vert k'\hat{\mathbf{z}}, \mu_1''',
\mu_2''' \rangle 
D^{1/2}_{\mu_1''' \mu_1'}(R) D^{1/2}_{\mu_2''' \mu_2'}(R)
\]
\beq 
\times 
\langle {1\over 2}, \mu_1', {1 \over 2}, \mu_2' \vert s', m_s' \rangle 
\langle l', m_l', s', m_s' \vert j \mu \rangle Y_{l'm_l'}
(\hat{\mathbf{z}}) D^{l'}_{m_l''' m_l'} (R).
\label{ap2.13}
\eeq
Using properties of Clebsch-Gordan coefficients (i.e. $D(R)<|> = <|>D(R)\otimes
D(R)$) we get 
\[
\langle  k, l, s \Vert  V^j \Vert k', l', s' \rangle
\]
\[ 
=\int d \hat{\mathbf{k}}'' d\hat{\mathbf{k}}' {1 \over 2j+1}
\sum_{\mu=-j}^j  D^{j*}_{\mu'' \mu} (R)
\langle {1\over 2}, \mu_1, {1 \over 2}, \mu_2 \vert s, m_s \rangle 
\langle l, m_l, s, m_s \vert j, \mu \rangle 
Y^*_{lm_l}( \hat{\mathbf{x}}\sin (\theta'')+\hat{\mathbf{z}}\cos(\theta'')  
\vert l,m_l)
\]
\[ 
\times  
\langle 
k''(\hat{\mathbf{x}}\sin (\theta'')+\hat{\mathbf{z}}\cos(\theta'') , 
\mu_1,\mu_2 \Vert V \Vert k'\hat{\mathbf{z}}, \mu_1',
\mu_2' \rangle 
\]
\[ 
\times 
\langle {1\over 2}, \mu_1', {1 \over 2}, \mu_2' \vert s', m_s' \rangle 
\langle l', m_l', s', m_s' \vert j, \mu''' \rangle Y_{l'm_L'}
(\hat{\mathbf{z}})  D^{j}_{\mu''' \mu} (R) 
\]
\[ 
=\int d \hat{\mathbf{k}}'' d\hat{\mathbf{k}}' {1 \over 2j+1}
\sum_{\mu''=-j}^j  
\langle {1\over 2}, \mu_1, {1 \over 2}, \mu_2 \vert s, m_s \rangle 
\langle l, m_l, s, m_s \vert j, \mu'' \rangle 
Y^*_{lm_l} ( \hat{\mathbf{x}}\sin (\theta'')+\hat{\mathbf{z}}\cos(\theta''))  
\]
\[ 
\times  
\langle 
k''(\hat{\mathbf{x}}\sin (\theta'')+\hat{\mathbf{z}}\cos(\theta'') , 
\mu_1,\mu_2 \Vert V \Vert k'\hat{\mathbf{z}}, \mu_1',
\mu_2' \rangle 
\]
\beq
\times 
\langle {1\over 2}, \mu_1', {1 \over 2}, \mu_2' \vert s', m_s' \rangle 
\langle l', m_l', s', m_s' \vert j, \mu'' \rangle 
Y_{l'm_l'} (\hat{\mathbf{z}}).  
\label{a.14}
\eeq
Since all of the dependence on $\phi',\theta',\phi''$ is in $R$,
and the integrand is independent of $R$, these three angular 
integrals can be computed giving the multiplicative
phase space factor of $8\pi^2$.  What remains is the following 
integral over the cosine of the angle between the final and
initial momentum: 
\[
\langle  k, l, s \Vert  V^j \Vert k', l', s' \rangle
\]
\[ 
=\int_{-1}^1 du''   {8 \pi^2 \over 2j+1}
\sum_{\mu''=-j}^j  
\langle {1\over 2}, \mu_1, {1 \over 2}, \mu_2 \vert s, m_s \rangle 
\langle l, m_l, s, m_s \vert j, \mu'' \rangle 
Y^*_{lm_l}( \hat{\mathbf{x}}\sqrt{1-u^{\prime\prime 2}})+\hat{\mathbf{z}}u'')  
\]
\[ 
\times  
\langle 
k''(\hat{\mathbf{x}}\sqrt{1-u^{\prime\prime 2}})+\hat{\mathbf{z}}u'') , 
\mu_1,\mu_2 \Vert V \Vert k'\hat{\mathbf{z}}), \mu_1',
\mu_2' \rangle 
\]
\beq 
\times 
\langle {1\over 2}, \mu_1', {1 \over 2}, \mu_2' \vert s', m_s' \rangle 
\langle l', m_l', s', m_s' \vert j, \mu'' \rangle 
Y_{l'm_l'}( \hat{\mathbf{z}}) .  
\label{ap2.15}
\eeq
The last thing that needs for be addressed for an explicit 
formula is the spherical harmonics
\beq
Y_{lm}^* 
( \hat{\mathbf{x}}\sqrt{1-u^{\prime\prime 2}})+\hat{\mathbf{z}}u'' ) 
=(-)^{m_l} \sqrt{{2l+1 \over 4 \pi}}
\sqrt{{(l-m_l')! \over (l+m_l)!}} P^m_l (u'')=
\sqrt{{2l+1 \over 4 \pi}}D^l_{m0}(R_y(\theta''))
\label{ap2.16}
\eeq
\beq
Y_{l'm_l'}( \hat{\mathbf{z}})  =
\delta_{m_l'0} \sqrt{{2l'+1 \over 4 \pi}}.
\label{ap2.17}
\eeq
Using these in the above expression we are left with a 1 dimensional integral
\[
\langle  k, l, s \Vert  V^j \Vert k', l', s' \rangle
\]
\[ 
=\int_{-1}^1 du'' {8 \pi^2 \over 2j+1}
\sum_{\mu''=-j}^j  
\langle {1\over 2}, \mu_1, {1 \over 2}, \mu_2 \vert s, m_s \rangle 
\langle l, m_l, s, m_s \vert j, \mu'' \rangle 
\sqrt{{2l+1 \over 4 \pi}}D^l_{m0}(R_y(\theta''))
\]
\[ 
\times  
\langle 
k''(\hat{\mathbf{x}}\sqrt{1-u^{\prime\prime 2}})+\hat{\mathbf{z}}u'') , 
\mu_1,\mu_2 \Vert V \Vert k'\hat{\mathbf{z}}, \mu_1',
\mu_2' \rangle 
\]
\beq 
\times 
\langle {1\over 2}, \mu_1', {1 \over 2}, \mu_2' \vert s', m_s' \rangle 
\langle l', 0, s', m_s' \vert j, \mu'' \rangle
\sqrt{{2l'+1 \over 4 \pi}} .
\label{ap2.17b}
\eeq
Cleaning this up gives the following expression for the partial 
wave amplitude:
\bigskip
\[
\langle  j , \mu , k, l, s \vert  V \vert j', \mu', k', l', s' \rangle =
\]
\[
\delta_{jj'} \delta_{\mu \mu'} 
\langle  k, l, s \Vert  V^j \Vert k', l', s' \rangle
\]
with 
\[
\langle  k, l, s \Vert  V^j \Vert k', l', s' \rangle
\]
\[ 
= {8 \pi^2 \over 2j+1}
\sqrt{{2l+1 \over 4 \pi}}
\sqrt{{2l'+1 \over 4 \pi}}
\sum_{\mu''=-j}^j \sum 
\langle {1\over 2}, \mu_1, {1 \over 2}, \mu_2 \vert s, m_s \rangle 
\langle l, m_l, s, m_s \vert j \mu'' \rangle 
\]
\[
\times
\int_{-1}^1 du''
D^l_{m0}(R_y(cos^{-1}(u''))
\langle 
k''(\hat{\mathbf{x}}\sqrt{1-u^{\prime\prime 2}})+\hat{\mathbf{z}}u'') , 
\mu_1,\mu_2 \Vert V \Vert k'\hat{\mathbf{z}}, \mu_1',
\mu_2' \rangle 
\]
\beq 
\times 
\langle {1\over 2}, \mu_1', {1 \over 2}, \mu_2' \vert s', m_s' \rangle 
\langle l', 0, s', m_s' \vert j \mu'' \rangle
\label{ap2.18}
\eeq
where all repeated spin indicies are summed.
This reduces the partial-wave integral to a one-dimensional integral.
In this case there are no traces, and no momentum-dependent spin
bases, but there are a number of spin sums.  

\bigskip
Explicit computation requires
\beq
D^l_{m0}(R_y(cos^{-1}(u'')) =
\sum_{s=0}^{2j} 
{\sqrt{j+m)!(j-m)!j!j!} \over
(j+m-s)!s!(s-m)!(j-s)!}
R_{11}^{j+m-s}
R_{12}^{s}
R_{21}^{s-m}
R_{22}^{j+s}
\label{ap2.19} 
\eeq
where all negative factorials are $^\infty$ and
\beq
R_{ij} (\cos^{-1}(u''))  =
\left (
\begin{array}{cc} 
\sqrt{{1+u^{\prime\prime 2}\over 2}} & \sqrt{{1-u^{\prime\prime 2}\over 2}} \\
-\sqrt{{1-u^{\prime\prime 2}\over 2}}& \sqrt{{1+u^{\prime\prime 2}\over 2}} \\
\end{array} 
\right ) 
\label{a.20}
\eeq
or 
\[
D^l_{m0}(R_y(cos^{-1}(u'')) =
\]
\beq
=\sum_{s=0}^{2j} 
{\sqrt{j+m)!(j-m)!j!j!} \over
(j+m-s)!s!(s-m)!(j-s)!}
(-1)^{s-m+0}
(\sqrt{{1+u^{\prime\prime 2}\over 2}})^{2j+m}
(\sqrt{{1-u^{\prime\prime 2}\over 2}})^{2s-m} .
\label{ap2.21}
\eeq

\section{Appendix 3}

The electromagnetic corrections to the nucleon-nucleon interaction have 
the following forms for the  
$pp$, $np$, and $nn$ systems:
\[
V_{em\, pp}(r) = 
\]
\beq
(v_{em,1}(r) + v_{em,2}(r) + v_{em,3}(r) + v_{em,4}(r)) I +
v_{em,6}(r) \pmb{\sigma}_1 \cdot \pmb{\sigma}_2 +
v_{em,9}(r) S_{12} + 
v_{em,12}(r) \mathbf{L} \cdot \mathbf{S}
\label{ap3.1}
\eeq

\beq
V_{em\, np}(r) = 
v_{em,5}(r) I +
v_{em,8}(r) \pmb{\sigma}_1 \cdot \pmb{\sigma}_2 +
v_{em,11}(r) S_{12} + v_{em,14}(r) \mathbf{L} \cdot \mathbf{S}
\label{ap3.2}
\eeq

\beq
V_{em\, nn}(r) = 
v_{em,7}(r) \pmb{\sigma}_1 \cdot \pmb{\sigma}_2 +
v_{em,10}(r) S_{12} + v_{em,13}(r) \mathbf{L} \cdot \mathbf{S}
\label{ap3.3}
\eeq

The Fourier transforms have the same structure as they do for 
the strong interactions.
\[
\tilde{V}_{em\, pp}(q) = 
(v_{em,1}(q) + v_{em,2}(q) + v_{em,3}(q) + v_{em,4}(q)) I +
\]
\beq
v_{em,6}(q) \pmb{\sigma}_1 \cdot \pmb{\sigma}_2 +
v_{em,9}(q) \tilde{S}_{12} + i v_{em,12}(q) (\mathbf{k}\times\mathbf{k}')\cdot 
\mathbf{S}.
\label{ap3.1b}
\eeq
\beq
\tilde{V}_{em\, np}(q) = 
v_{em,5}(q) I +
v_{em,8}(q) \pmb{\sigma}_1 \cdot \pmb{\sigma}_2 +
v_{em,11}(q) \tilde{S}_{12} + 
 i v_{em,14}(q) (\mathbf{k}\times\mathbf{k}')\cdot \mathbf{S}.
\label{ap3.2b}
\eeq

\beq
\tilde{V}_{em\, nn}(q) = 
v_{em,7}(q) \pmb{\sigma}_1 \cdot \pmb{\sigma}_2 +
v_{em,10}(q) \tilde{S}_{12} + i v_{em,13}(q) (\mathbf{k}\times\mathbf{k}')\cdot 
\mathbf{S}.
\label{ap3.3b}
\eeq
where
\beq
\tilde{S}_{12} :=
-3((\mathbf{q}\cdot \pmb{\sigma}_1)
(\mathbf{q}\cdot \pmb{\sigma}_2)- \pmb{\sigma}_1 \cdot \pmb{\sigma}_2).
\eeq

The coefficients of
$I,\pmb{\sigma}_1 \cdot \pmb{\sigma}_2$ are
\beq
\tilde{v}_{em,n}(q) =  
{1 \over 2 \pi^2 }\int_0^\infty j_0 (qr) v_{em,n}(r) r^2 dr
\qquad n \in \{1,2,3,4,5,6,7,8 \};
\label{ap3.4}
\eeq
the coefficients of $i (\mathbf{k}\times\mathbf{k}')\cdot \mathbf{S}$ are 
\beq
\tilde{v}_{em,n}(q) = {1 \over 2 \pi^2 q }\int_0^\infty j_1 (qr) v_{em,n}(r) 
r^3 dr
\qquad n \in \{12,13,14 \}
\label{ap3.5}
\eeq
and the coefficients of $\tilde{S}_{12}$,
are
\beq
\tilde{v}_{em,n}(q) =  {1 \over 2 \pi^2 q^2}\int_0^\infty j_2 (qr) v_{em,n}(r) r^2 dr
\qquad n \in \{9,10,11 \}
\label{ap3.6}
\eeq
where the individual terms are
\beq
v_{em,1}(r) = 
{\alpha \hbar c  \over r}(1 - (1+{11\over 16} br +{3\over 16} (br)^2 
+{1\over 48} (br)^3
)e^{-br}),
\label{ap3.7}
\eeq
\beq
v_{em,2}(r) = -{\alpha (\hbar c)^3 \over 4 m_p^2}
{b^3 \over 16} (1+ br +{1\over 3} (br)^2 )e^{-br}),
\label{ap3.8}
\eeq

\beq
v_{em,3}(r) = - {1 \over m_p}
v_{em}(1)^2,
\label{ap3.9}
\eeq

\beq
v_{em,4}(r) =
{2 \alpha \over 3 \pi}(-\gamma - {5 \over 6} + \vert \ln (a r) \vert)  
v_{em,1}(r),
\label{ap3.10}
\eeq

\beq
v_{em,5}(r)= \alpha \hbar c \beta 
{b^3 \over 384} (15+ 15 br +6  (br)^2 + (br)^3 )e^{-br}),
\label{ap3.11}
\eeq

\beq
v_{em,6}(r)= - {\alpha (\hbar c)^3 \mu_p^2  \over 6 m_p^2} 
{b^3 \over 16} (1+ br +{1\over 3} (br)^2 )e^{-br}),
\label{ap3.12}
\eeq

\beq
v_{em,7}(r) =  - {\alpha (\hbar c)^3 \mu_n^2  \over 6 m_n^2} 
{b^3 \over 16} (1+ br +{1\over 3} (br)^2 )e^{-br}),
\label{ap3.13}
\eeq

\beq
v_{em,8}(r) =  - {\alpha (\hbar c)^3 \mu_n \mu_p  \over 6 m_n m_p} 
{b^3 \over 16} (1+ br +{1\over 3} (br)^2 )e^{-br}),
\label{ap3.14}
\eeq

\beq
v_{em,9}(r) =  - {\alpha (\hbar c)^3 \mu_p^2  \over 4 m_p^2} 
{1 \over r^3} ( 1 -(1+ br +{1\over 2} (br)^2 
+{1\over 6} (br)^3 +{1\over 24} (br)^4
+{1\over 144} (br)^5 )e^{-br}),
\label{ap3.15}
\eeq

\beq
v_{em,10}(r) =  - {\alpha (\hbar c)^3 \mu_n^2  \over 4 m_n^2} 
{1 \over r^3} ( 1 -(1+ br +{1\over 2} (br)^2 
+{1\over 6} (br)^3 +{1\over 24} (br)^4
+{1\over 144} (br)^5 )e^{-br}),
\label{ap3.16}
\eeq

\beq
v_{em,11}(r) =  - {\alpha (\hbar c)^3 \mu_p \mu_n  \over 4 m_p m_n} 
{1 \over r^3} ( 1 -(1+ br +{1\over 2} (br)^2 
+{1\over 6} (br)^3 +{1\over 24} (br)^4
+{1\over 144} (br)^5 )e^{-br}),
\label{ap3.17}
\eeq

\beq
v_{em,12}(r) =  - {\alpha (\hbar c)^3 (4 \mu_p-1)\mu_p \mu_n  \over 2 m_p^2} 
{1 \over r^3} ( 1 -(1+ br +{1\over 2} (br)^2 
+{7\over 48} (br)^3  )e^{-br}),
\label{ap3.18}
\eeq

\beq
v_{em,13}(r)=0,
\label{ap3.19}
\eeq
and
\beq
v_{em,14}(r) =  - {\alpha (\hbar c)^3 \mu_n  \over 2 m_n m_r}
{1 \over r^3} ( 1 -(1+ br +{1\over 2} (br)^2 
+{7\over 48} (br)^3  )e^{-br})
\label{ap3.20}
\eeq
where $b=4.27$ and $a=m_e/(\hbar c)$, $\mu_p=2.7928474$, and 
$\mu_n=-1.9130427$, $\beta=.0189$.

The Fourier transform of most of the terms in the potential can be 
computed using direct integration, the identities
\beq
j_1(x) = - {d \over dx} j_0(x)
\qquad
{1 \over x} j_2(x) = - {d \over dx} {j_1(x)\over x} ,
\label{ap3.21}
\eeq
and the following relation \cite{gradshteyn1},
%
%
with  $\nu=l+{1 \over 2}$ and $\mu =n +1/2$, gives the relation
\[
\int_0^\infty e^{-b r} j_{l} (qr) r^{n} dx
\]

\beq
={q^l \over b^{n+l+1}}{(n+l)! \over (2l+1)!!}
F({n+l+1 \over 2},{n+l+2 \over 2}, {2l+3 \over 2},-q^2/b^2)
\label{ap3.24}
\eeq
which is valid for $n+l>-1$.

The only integral that can not be computed using these formulas involves
the $\vert \ln(k r) \vert$ term that appears in $v_{em\, 4}(r)$, which
is an approximation to the vacuum polarization correction to the $pp$
interaction.   The required integrals, which are evaluated 
in this appendix are:

\beq
\int_0^\infty j_0(qr) \vert \ln (ar) \vert  r dr =
{a^2 \over q} [\gamma + 3\ln (q/a) - 2\mbox{ci}(q/a)],
\label{ap3.33}
\eeq

\[
\int_0^\infty j_0(qr) \vert \ln (ar) \vert r e^{-br } dr =
\]
\[
{a^3 \over b^2 + q^2}[b \tan^{-1}({q \over b})-q\gamma
+ {q \over 2} \ln (q^2/a^2 + b^2/a^2)] +
\]
\beq
{a^3 \over q^2 + b^2} 
\left [q {(E_1({b-iq \over a}) + E_1({b+iq \over a})) \over 2}+
b {(E_1({b-iq \over a}) - E_1({b+iq \over a})) \over 2i}\right ],
\label{ap3.34}
\eeq

\beq
\int_0^\infty j_0(qr) \vert \ln (ar) \vert r^2 e^{-br } dr =
-{\partial \over \partial b}
\int_0^\infty j_0(qr) \vert \ln (ar) \vert r e^{-br } dr ,
\label{ap3.36}
\eeq

\beq
\int_0^\infty j_0(qr) \vert \ln (ar) \vert r^3 e^{-br } dr =
{\partial^2 \over \partial b^2} \int_0^\infty j_0(qr) \vert \ln (ar) \vert r e^{-br } dr ,
\label{ap3.37}
\eeq

\beq
\int_0^\infty j_0(qr) \vert \ln (ar) \vert r^4 e^{-br } dr =
-{\partial^3 \over \partial b^3}
\int_0^\infty j_0(qr) \vert \ln (ar) \vert r e^{-br } dr
\label{ap3.38}
\eeq
and
\beq
\int_0^\infty j_2(qr) r^{-1}  dr = \lim_{x \to 0} j_1(x)/x = 1/3 .
\label{ap3.39}
\eeq

The vacuum polarization integral only appears in the Coulomb potential
which has the approximate form given in \cite{ak} - this
approximation is adequate for binding energy calculations.  It appears in
the following contribution to the proton-proton interaction
\beq
v_{em(4)}(r) =
{2 \alpha \over 3 \pi} (-\gamma - {5 \over 6} + \vert \ln (a r) \vert + 
{6\pi k r \over 8})
{\alpha \hbar c \over r} (1 - e^{-br}(1+{11 \over 16} br +{3 \over 16} (br)^2  
+{1 \over 48} (br)^3 ). 
\label{ap3.52}
\eeq

The Fourier Bessel transform of this interaction,
\[
{1 \over 2 \pi^2 q}\int_0^{\infty}  j_0(qr) V_{em(4)}(r) r^2 dr =
\]
\beq
{1 \over 2 \pi^2 q^2} \int_0^\infty \sin (qr) 
{2 \alpha \over 3 \pi}(-\gamma - {5 \over 6} + \vert \ln (a r) \vert + 
{6\pi k r \over 8})
{\alpha \hbar c } (1 - e^{-br}(1+{11 \over 16} br +{3 \over 16} (br)^2  
+{1 \over 48} (br)^3 ), 
dr
\label{ap3.53}
\eeq

can be computed analytically.  There are three types of contributions
\beq
(I) =
{1 \over 2 \pi^2 q^2} \int_0^\infty \sin (qr) 
{2 \alpha \over 3 \pi}(-\gamma - {5 \over 6} )
{\alpha \hbar c } (1 - e^{-br}(1+{11 \over 16} br +{3 \over 16} (br)^2  
+{1 \over 48} (br)^3 ) 
dr,
\label{ap3.54}
\eeq
\beq
(II)= 
{1 \over 2 \pi^2 q^2} \int_0^\infty \sin (qr) 
{2 \alpha \over 3 \pi}\vert \ln (a r) \vert 
{\alpha \hbar c } (1 - e^{-br}(1+{11 \over 16} br +{3 \over 16} (br)^2  
+{1 \over 48} (br)^3 ), 
dr,
\label{ap3.55}
\eeq
and
\beq
(III) = 
{1 \over 2 \pi^2 q^2} \int_0^\infty \sin (qr) 
{2 \alpha \over 3 \pi} 
{6\pi k r \over 8}
{\alpha \hbar c } (1 - e^{-br}(1+{11 \over 16} br +{3 \over 16} (br)^2  
+{1 \over 48} (br)^3 ) 
dr .
\label{ap3.56}
\eeq
Integrals of the form (I) and (III) have the same form as the integrals 
discussed above.  To calculate the integral (II)  
first replace $r'=ar \to r = r'/a$ to get
\[
(II)= 
{1 \over 2 a \pi^2 q^2} \int_0^\infty \sin (qr'/a) 
{2 \alpha \over 3 \pi}\vert \ln (r') \vert 
{\alpha \hbar c } (1 - e^{-br'/a}(1+{11 \over 16} {br'\over a} +{3 \over 16} 
({br'\over a})^2  
+{1 \over 48} ({br' \over a})^3 )) 
dr' =
\]
\[
= {1 \over 2 a \pi^2 q^2} 
{2 \alpha \over 3 \pi}
{\alpha \hbar c }
\int_0^\infty \sin (qr'/a) 
\vert \ln (r') \vert 
dr'
\]
\beq
- {1 \over 2 a \pi^2 q^2}{2 \alpha \over 3 \pi}
{\alpha \hbar c }  (1+{11 \over 16} (-b {d \over db}) +{3 \over 16} 
(b^2 {d^2 \over d^2b})  
+{1 \over 48} (-b^3 {d^3 \over d^3b}) )
\int_0^\infty \sin (qr'/a) e^{-br'/a}
\vert \ln (r') \vert 
dr'.
\label{ap3.57}
\eeq

Two integrals need to be performed to compute this term.  They are
\beq
\int_0^\infty \sin (qr'/a) 
\vert \ln (r') \vert 
dr' 
\label{ap3.58}
\eeq
and 
\beq
\int_0^\infty \sin (qr'/a) e^{-br'/a} 
\vert \ln (r') \vert 
dr'.
\label{ap3.59}
\eeq

The integral 
\[
\int_0^\infty \sin (qr'/a) 
\vert \ln (r') \vert  
dr' = 
\]
\beq
-\int_0^1 \sin (qr'/a) 
\ln (r')  
dr' +
\lim_{d \to 0}[ \int_0^\infty \sin (qr'/a)e^{-d r'} 
\ln (r')dr' - \int_0^1 \sin (qr'/a) 
\ln (r')dr'].  
\label{ap3.60}
\eeq
These integrals can be found in \cite{gradshteyn2}:
\beq
\int_0^1  \sin(qx) \ln (x)  dx = -{1 \over q} [\gamma + \ln (q) - \mbox{ci}(q)]
\label{ap3.61}
\eeq
where 
\beq
\mbox{ci}(x) = \mbox{Ci}(x)= -\int_x^\infty {\cos(t) \over t}
\label{ap3.62}
\eeq
and 
\beq
\int_0^\infty e^{- br} \sin(qr) \ln (r) dr =
{1 \over b^2 + q^2}[b \tan^{-1}({q \over b})-q\gamma 
+ {q \over 2} \ln (q^2 + b^2)].
\label{ap3.63}
\eeq
The quantity $\gamma$ is the Euler constant.

Thus the first of the required integrals needed to compute the 
vacuum polarization contribution is 
\[
\int_0^\infty \vert \ln (r') \vert \sin(qr'/a) dr' = 
\]
\[
-\int_0^1   \sin(qr'/a) \ln (r') dr' + 
\lim_{d \to 0} [\int_0^\infty e^{- dr'} \sin(qr'/a) \ln (r') dr' 
-\int_0^1  \ln (r')  \sin(qr'/a)] =
\]
\[
-2 \int_0^1 \ln (r')  \sin(qr'/a) dr' + 
\lim_{d \to 0} \int_0^\infty e^{- dr'} \sin(qr'/a) \ln (r') dr' 
=
\]
\[
{2a \over q} [\gamma + \ln (q/a) - \mbox{ci}(q/a)]
+
{a \over q}[-\gamma  
+ {1 \over 2} \ln (q^2/a^2)] =
\]
\beq
{a \over q} [\gamma + 3\ln (q/a) - 2\mbox{ci}(q/a)] .
\label{ap3.64}
\eeq
Returning to the original expression - the first term in (II) 
is 
\[
{1 \over 2 a \pi^2 q^2} 
{2 \alpha \over 3 \pi}
{\alpha \hbar c }
\int_0^\infty \sin (qr'/a) 
\vert \ln (r') \vert 
dr'
\]
\beq
={\alpha^2 \hbar c \over 3  \pi^3 q^3} 
[\gamma + 3\ln (q/a) - 2\mbox{ci}(q/a)]
\label{ap3.65}
\eeq
where $\gamma$ is the Euler constant.
Or
\beq
{1 \over 2 \pi^2 q^2} \int_0^\infty \sin (qr) 
{2 \alpha \over 3 \pi}\vert \ln (a r) \vert 
{\alpha \hbar c } 
={\alpha^2 \hbar c \over 3 \pi^3 q^3} 
[\gamma + 3\ln (q/a) - 2\mbox{ci}(q/a)].
\label{ap3.66}
\eeq
We also have to compute the second term in (II).  The integral 
that is needed is
\beq
-{\alpha^2 \hbar c \over 3 \pi^3 aq^2} 
 \int_0^{ \infty} \sin(qr'/a) \vert \ln (r') \vert 
e^{-br'/a}(1+{11 \over 16} {br'\over a} +{3 \over 16} 
({br'\over a})^2  
+{1 \over 48} ({br' \over a})^3 ) 
dr' .
\label{ap3.67}
\eeq
This can be evaluated using 
\[ 
- \int_0^{ \infty} \sin(qr'/a) \vert \ln (r') \vert 
e^{-br'/a} dr' =
\]
\beq
\int_0^{ \infty}  \sin(qr'/a) \ln (r')  
e^{-br'/a}  dr' - 2  \int_1^{ \infty}
 \sin(qr'/a) \ln (r')  
e^{-br'/a}  dr'
\label{ap3.68}
\eeq
by differentiation with respect to $b$. The integral
\[
\int_1^{ \infty}
\sin(qr'/a) \ln (r') e^{-br'/a} dr' 
  =
\]
\[
{1 \over 2i} 
\int_1^{ \infty}
 ( e^{(iq/a-b/a)r'}-e^{(-iq/a-b/a)r'}) \ln (r') dr' =
\]
\[
- {a \over 2i} 
\int_1^{ \infty} {dr'\over  r'}
[{ e^{(iq/a-b/a)r'} \over iq-b}+ { e^{(-iq/a-b/a)r'} \over iq+b}]
\]
\[
- {a \over 2i}
[{E_1({b-iq \over a}) \over iq-b}+ {E_1({b+iq \over a}) \over iq+b}]
\]
\[
{a \over 2i (q^2 + b^2)} [
(iq +b)E_1({b-iq \over a}) + (iq-b)E_1({b+iq \over a}] =
\]
\beq
{a \over q^2 + b^2} [
\left [q {(E_1({b-iq \over a}) + E_1({b+iq \over a})) \over 2}+
b {(E_1({b-iq \over a}) - E_1({b+iq \over a})) \over 2i}\right ] .
\label{ap3.69}
\eeq
Thus
\[
\int_1^{ \infty}
\sin(qr'/a) \ln (r') e^{-br'/a} dr' 
\]
\beq
={a \over q^2 + b^2} [
\left [q {(E_1({b-iq \over a}) + E_1({b+iq \over a})) \over 2}+
b {(E_1({b-iq \over a}) - E_1({b+iq \over a})) \over 2i}\right ].
\label{ap3.70}
\eeq
Combining the two integrals gives
\[
-{\alpha^2 \hbar c \over 3 \pi^3 aq^2} 
 \int_0^{ \infty} \sin(qr'/a) \vert \ln (r') \vert 
e^{-br'/a} dr' =
\]
\[
-{\alpha^2 \hbar c \over 3 \pi^3 aq^2} \times \left \{ \left . \right . 
\right .
\]
\[
{a \over b^2 + q^2}[b \tan^{-1}({q \over b})-q\gamma
+ {q \over 2} \ln (q^2/a^2 + b^2/a^2)] +
\]
\beq
\left .
{a \over q^2 + b^2} 
\left [q {(E_1({b-iq \over a}) + E_1({b+iq \over a})) \over 2}+
b {(E_1({b-iq \over a}) - E_1({b+iq \over a})) \over 2i}\right ]
\right \} .
\label{ap3.71}
\eeq
Putting all of the parts together gives the $ln (r)$  contribution to 
the vacuum polarization 
contribution
\[
(II)_b = 
{\alpha^2 \hbar c \over 3  \pi^3 q^3} 
[\gamma + 3\ln (q/a) - 2\mbox{ci}(q/a)] + 
\]
\[
-(1+{11 \over 16} (-b {d \over db}) +{3 \over 16} 
(b^2 {d^2 \over db^2})  
+{1 \over 48} (-b^3 {d^3 \over db^3}) )
\]
\[
{\alpha^2 \hbar c \over 3 \pi^3 aq^2} \times \left \{ \right .
\]
\[
{a \over b^2 + q^2}[b \tan^{-1}({q \over b})-q\gamma 
+ {q \over 2} \ln (q^2/a^2 + b^2/a^2)] +
\]
\beq
\left .
{a \over q^2 + b^2} [
\left [q {(E_1({b-iq \over a}) + E_1({b+iq \over a})) \over 2}+
b {(E_1({b-iq \over a}) - E_1({b+iq \over a})) \over 2i}\right ]
\right \} .
\label{ap3.72}
\eeq
This needs to be added to (I) and (II) to get the full vacuum 
polarization integral.
These integrals can be computed using the methods used for all of the
other potentials.

The exponential integrals have simple derivatives
\beq
{d \over dx} E_1(x)= - {e^{-x} \over x} .
\label{ap3.73}
\eeq

\clearpage

\vfill\eject

\begin{figure}
\begin{center}
\includegraphics[width=15.0cm,clip]{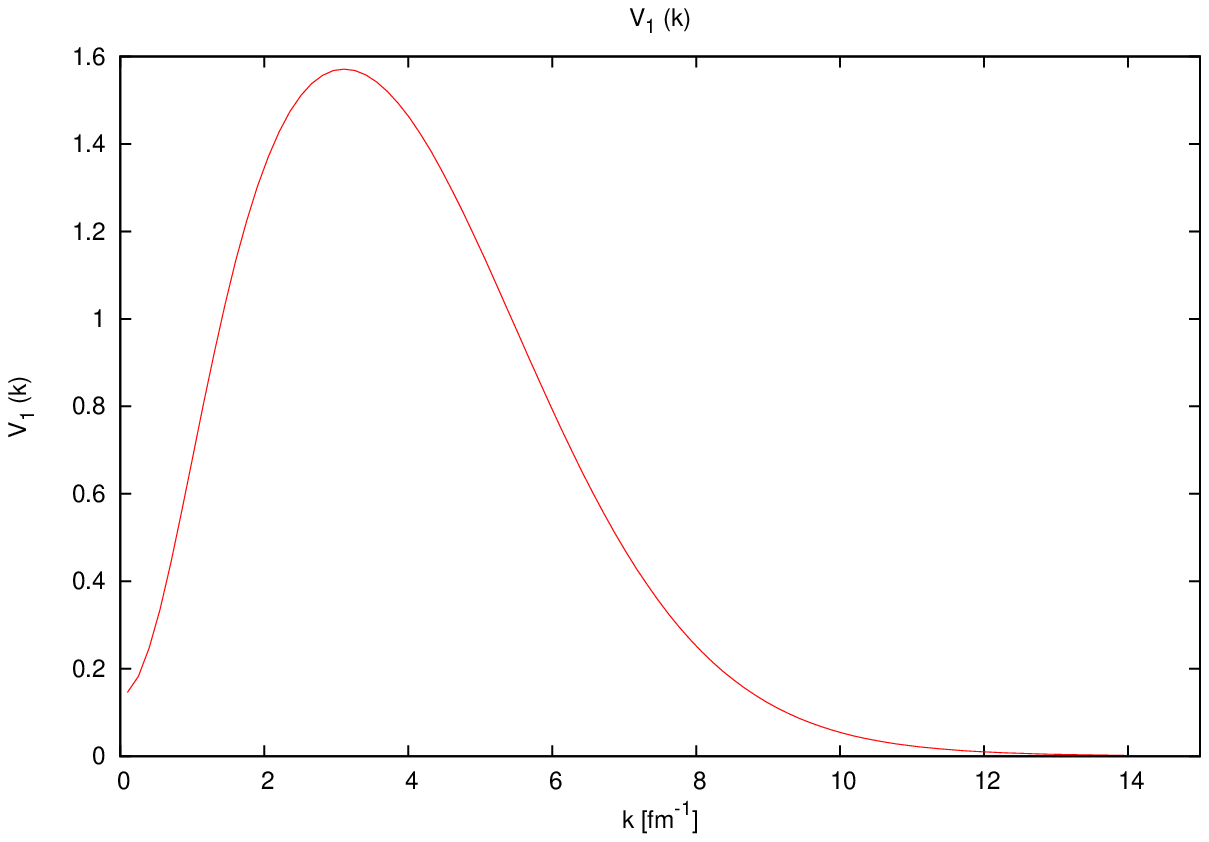}
\caption[Short caption for figure 1]{\label{labelFig1} 
$V_{1}(k)$. 
}
\end{center}
\label{fig.1}
\end{figure}

\begin{figure}
\begin{center}
\includegraphics[width=15.0cm,clip]{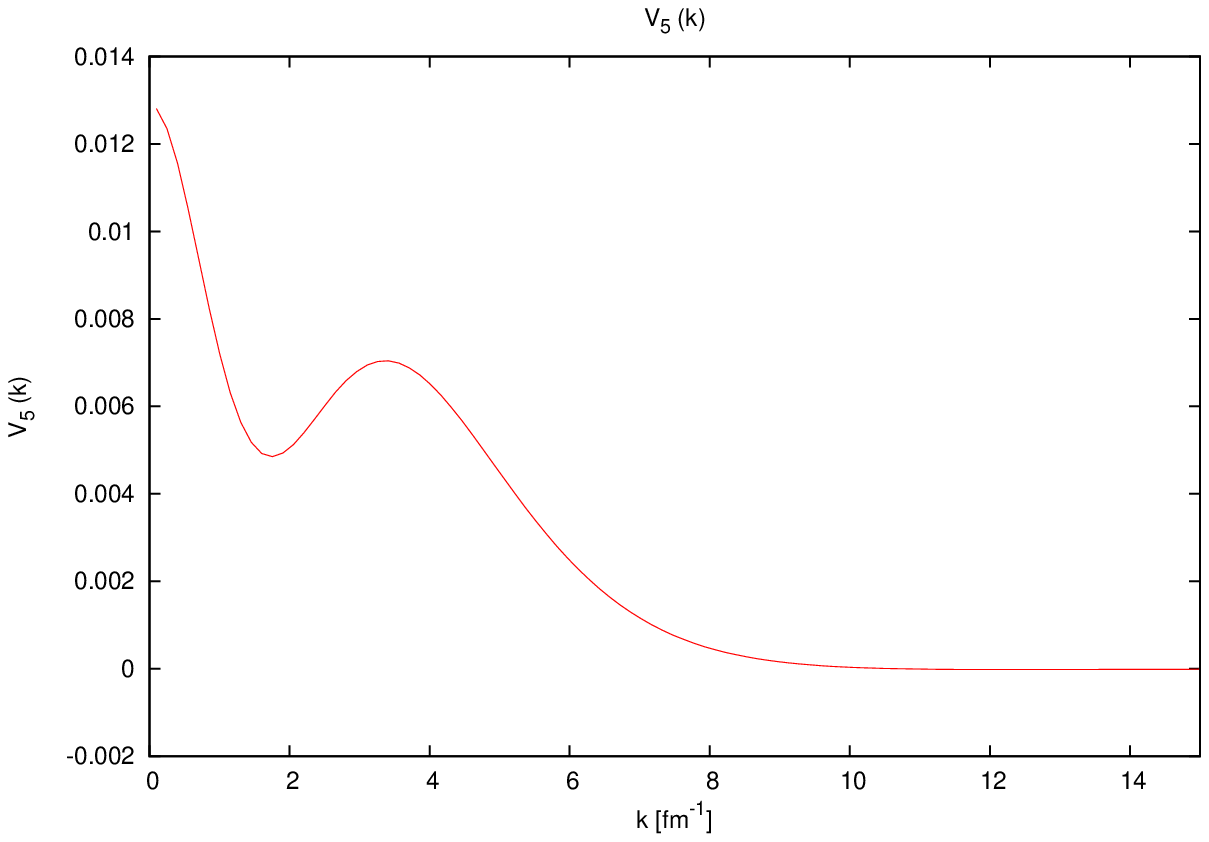}
\caption[Short caption for figure 2]{\label{labelFig2} 
$V_{5}(k)$. 
}
\end{center}
\label{fig.2}
\end{figure}

\begin{figure}
\begin{center}
\includegraphics[width=15.0cm,clip]{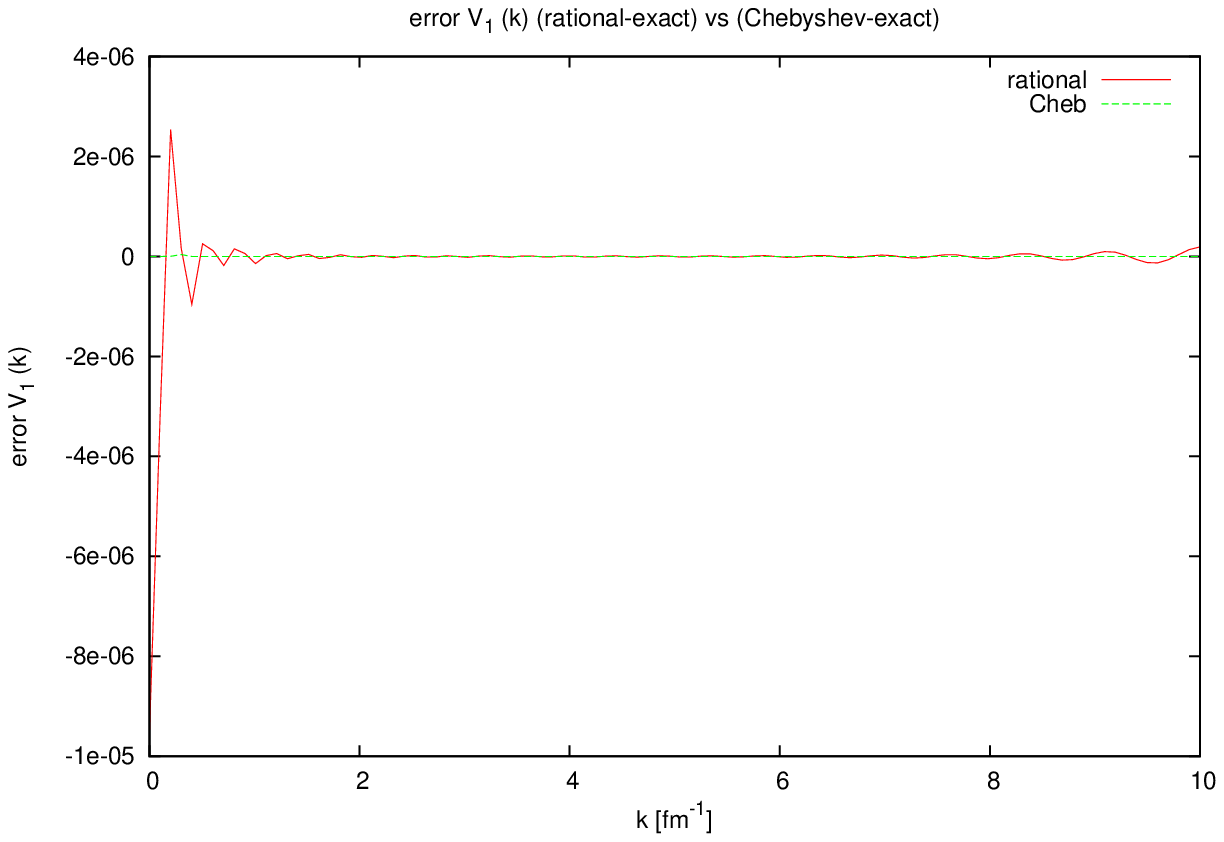}
\caption[Short caption for figure 3]{\label{labelFig3} 
$\Delta V_{1\, rational}(k) ,\,
\Delta V_{1\, Chebyshev}(k)$. 
}
\end{center}
\label{fig.3}
\end{figure}

\begin{figure}
\begin{center}
\includegraphics[width=15.0cm,clip]{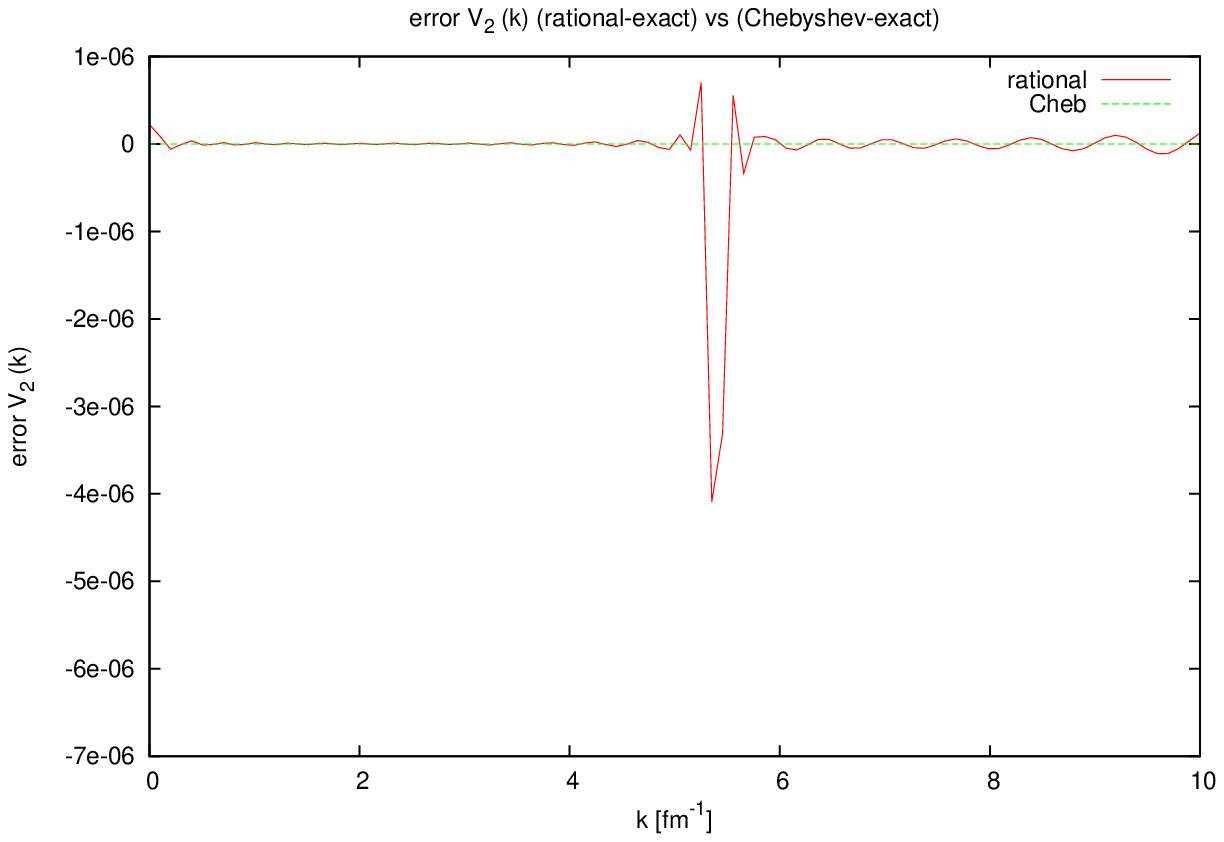}
\caption[Short caption for figure 4]{\label{labelFig4}
$\Delta V_{2\, rational}(k) ,\,
\Delta V_{2\, Chebyshev}(k) $.
}
\end{center}
\label{fig.4}
\end{figure}

\begin{figure}
\begin{center}
\includegraphics[width=15.0cm,clip]{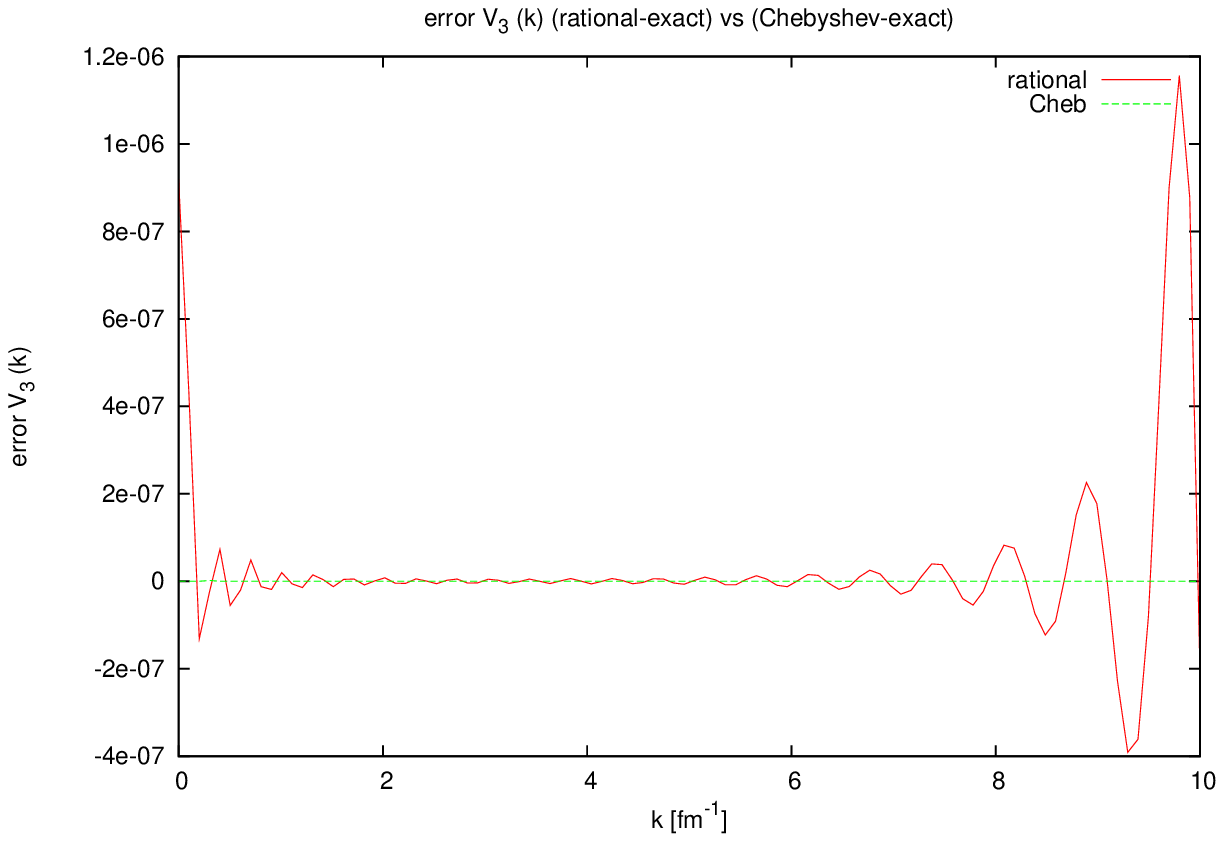}
\caption[Short caption for figure 5]{\label{labelFig5} 
$\Delta V_{3\, rational}(k),\, 
\Delta V_{3\, Chebyshev}(k)$. 
}
\end{center}
\label{fig.5}
\end{figure}

\begin{figure}
\begin{center}
\includegraphics[width=15.0cm,clip]{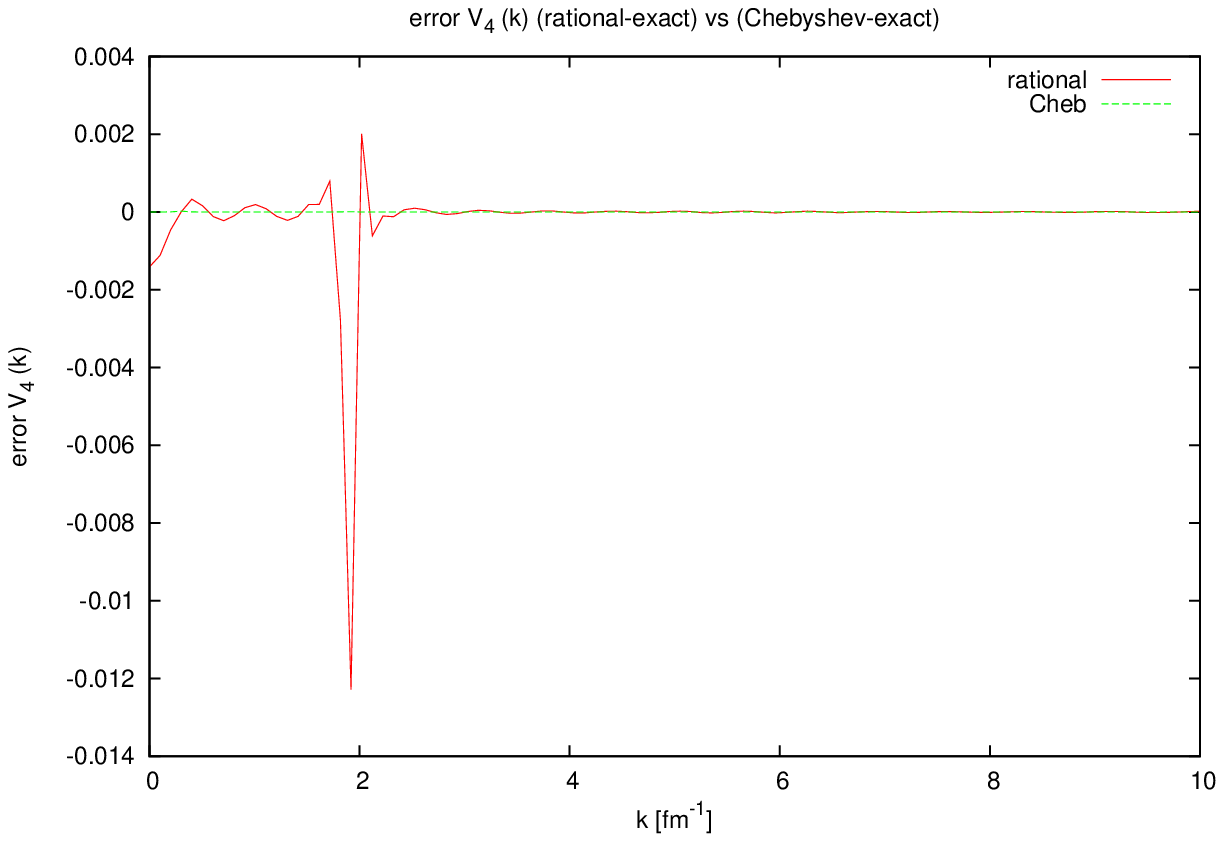}
\caption[Short caption for figure 6]{\label{labelFig6}
$\Delta V_{4\, rational}(k),\, 
\Delta V_{4\, Chebyshev}(k)$.
}
\end{center}
\label{fig.6}
\end{figure}

\begin{figure}
\begin{center}
\includegraphics[width=15.0cm,clip]{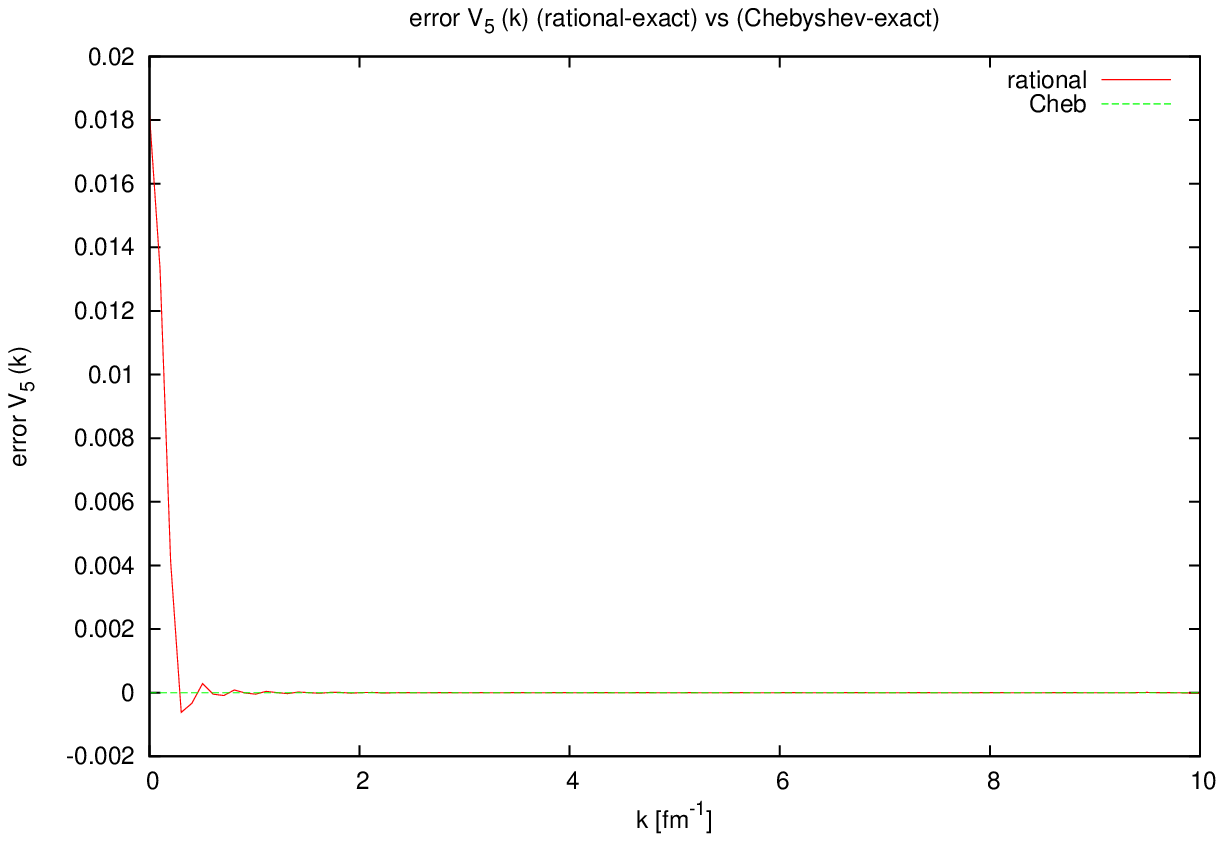}
\caption[Short caption for figure 7]{\label{labelFig7} 
$\Delta V_{5\, rational}(k),\, 
\Delta V_{5\, Chebyshev}(k)$. 
}
\end{center}
\label{fig.7}
\end{figure}

\begin{figure}
\begin{center}
\includegraphics[width=15.0cm,clip]{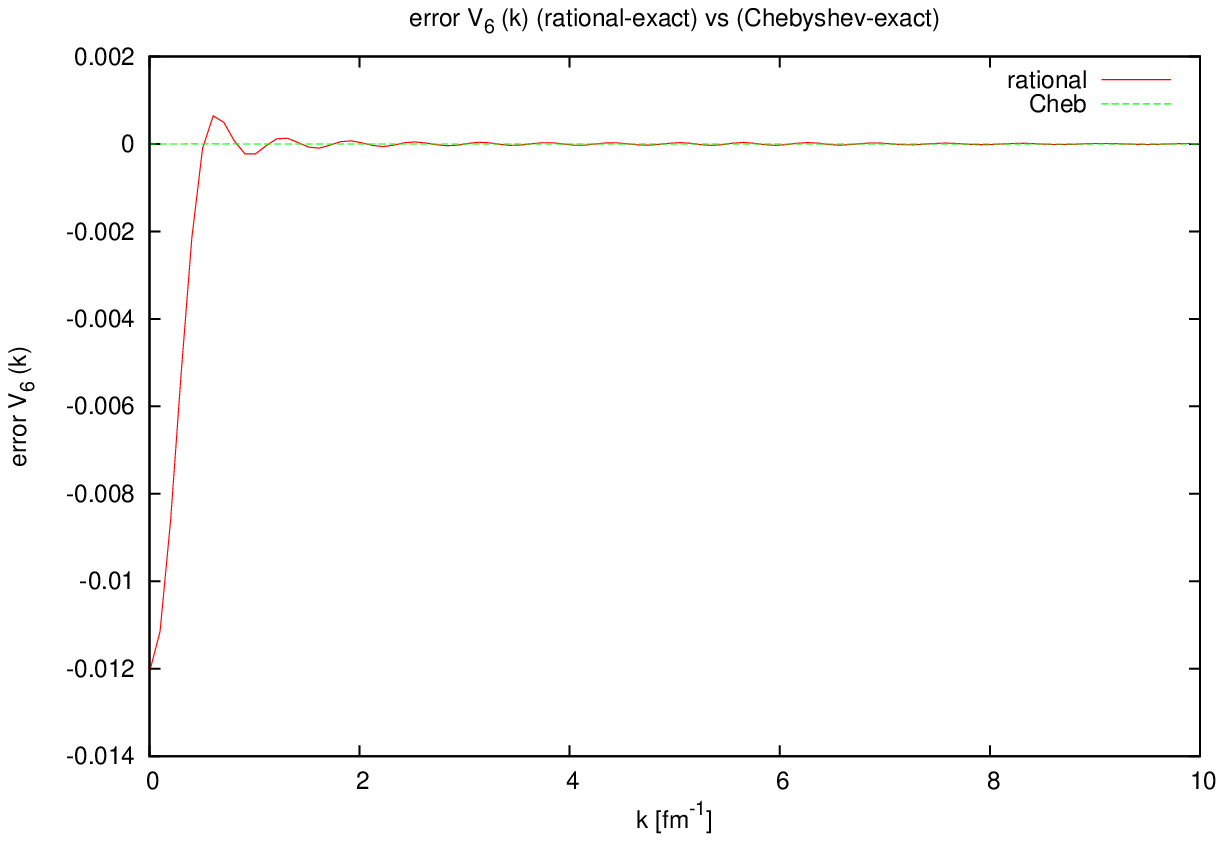}
\caption[Short caption for figure 8]{\label{labelFig8}
$\Delta V_{6\, rational}(k),\,
\Delta V_{6\, Chebyshev}(k)$.
}
\end{center}
\label{fig.8}
\end{figure}

\begin{figure}
\begin{center}
\includegraphics[width=15.0cm,clip]{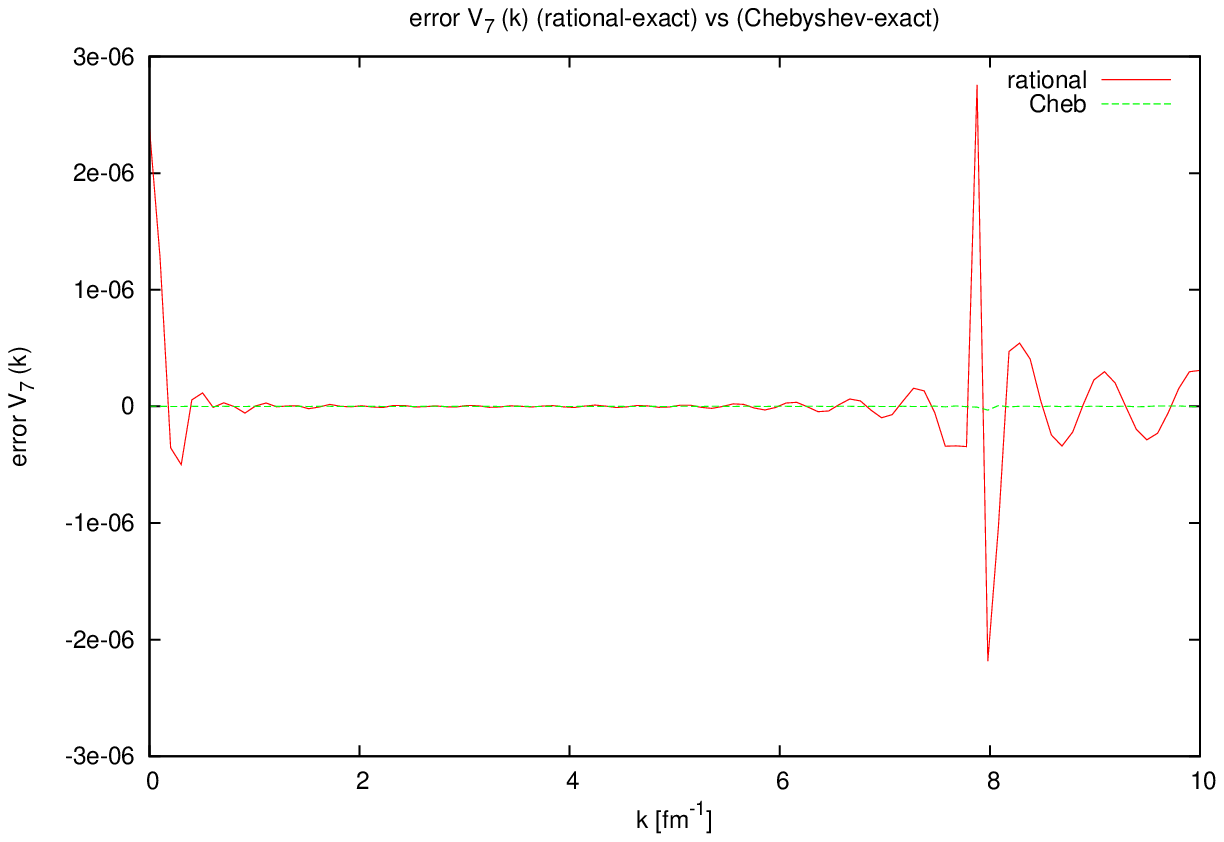}
\caption[Short caption for figure 9]{\label{labelFig9} 
$\Delta V_{7\, rational}(k),\,
\Delta V_{7\, Chebyshev}(k)$. 
}
\end{center}
\label{fig.9}
\end{figure}

\begin{figure}
\begin{center}
\includegraphics[width=15.0cm,clip]{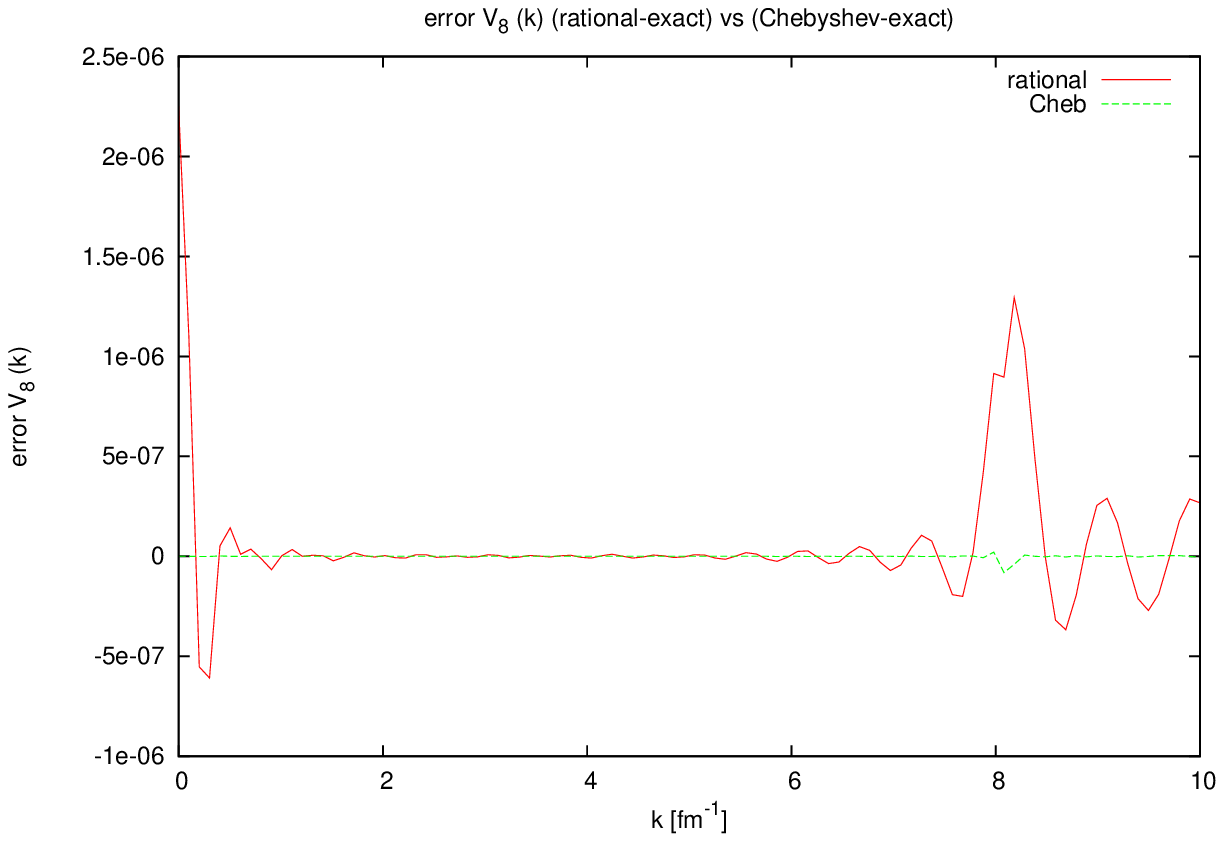}
\caption[Short caption for figure 10]{\label{labelFig10}
$\Delta V_{8\, rational}(k),\,
\Delta V_{8\, Chebyshev}(k)$.
}
\end{center}
\label{fig.10}
\end{figure}

\begin{figure}
\begin{center}
\includegraphics[width=15.0cm,clip]{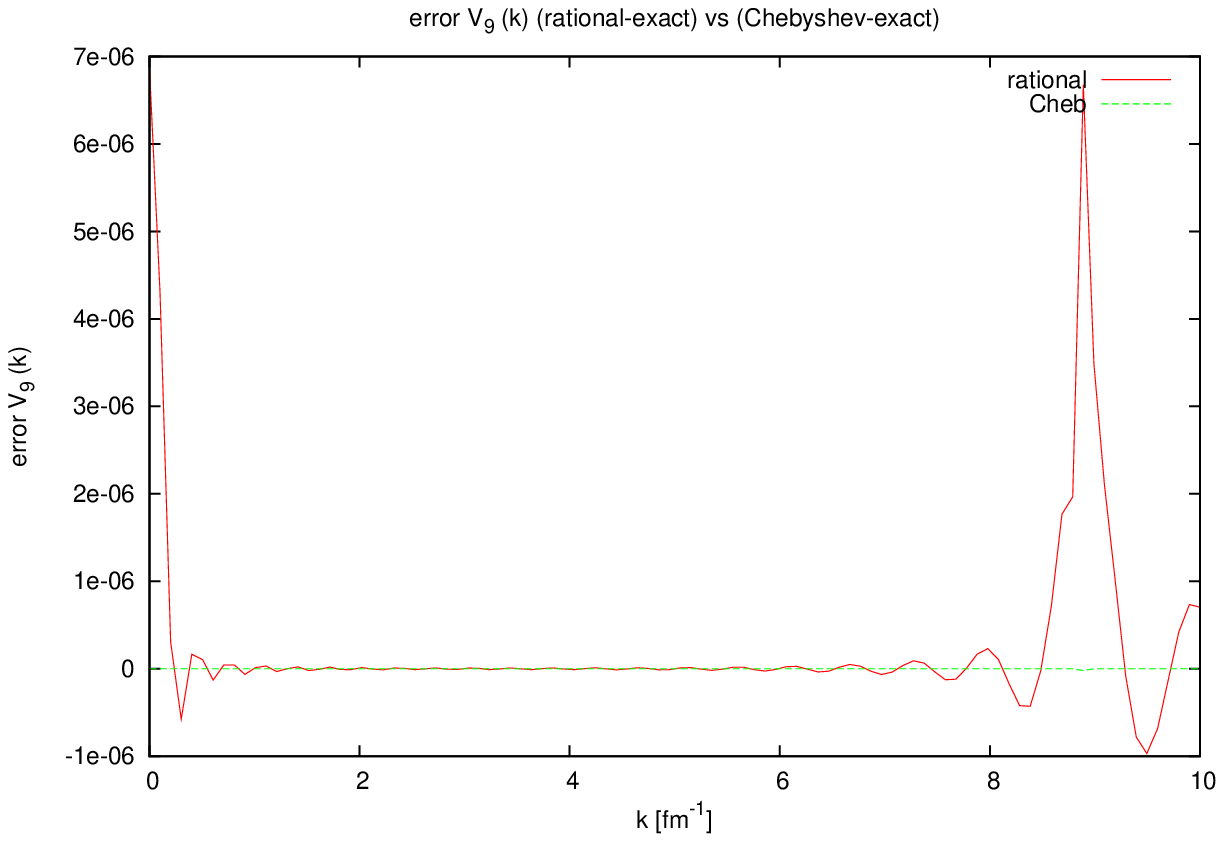}
\caption[Short caption for figure 11]{\label{labelFig11} 
$\Delta V_{9b\, rational}(k),\,
\Delta V_{9b\, Chebyshev}(k)$. 
}
\end{center}
\label{fig.11}
\end{figure}

\begin{figure}
\begin{center}
\includegraphics[width=15.0cm,clip]{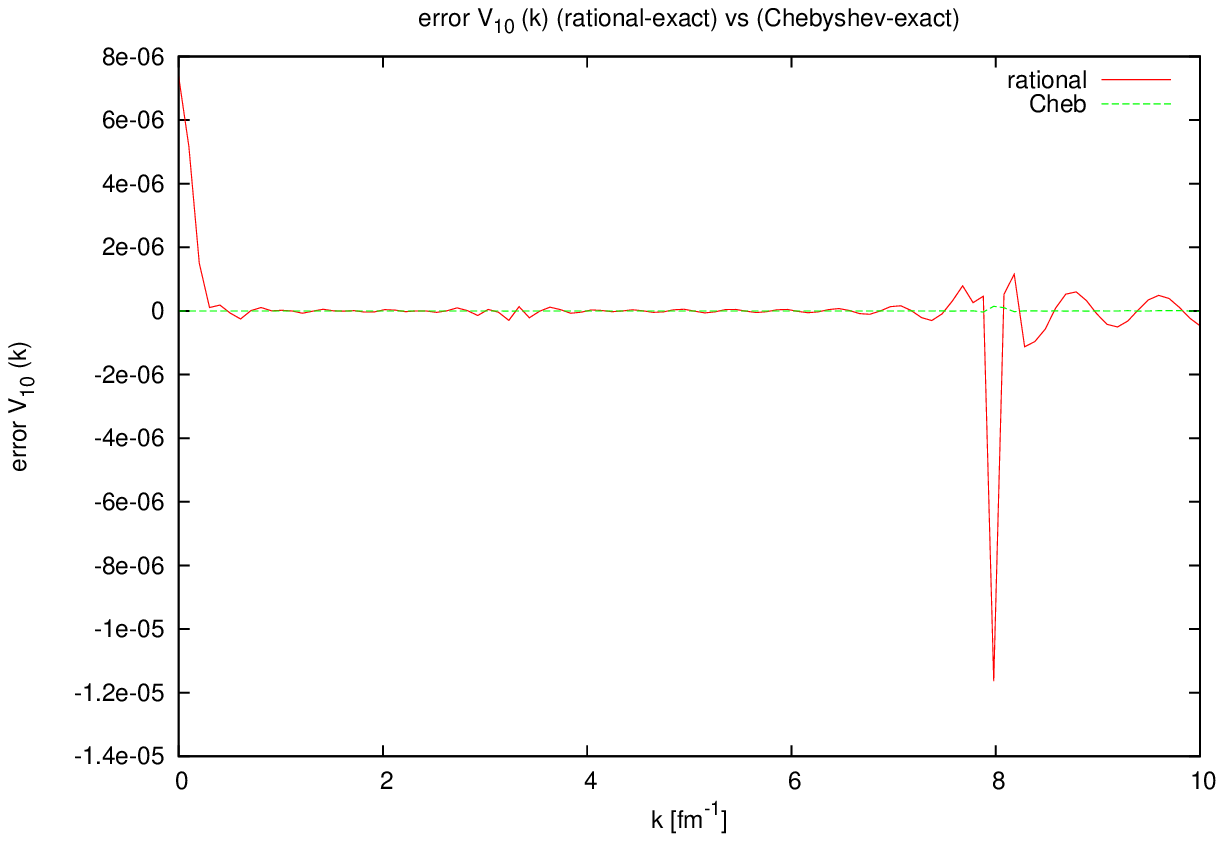}
\caption[Short caption for figure 12]{\label{labelFig12}
$\Delta V_{10b\, rational}(k),\,
\Delta V_{10b\, Chebyshev}(k)$.
}
\end{center}
\label{fig.12}
\end{figure}

\begin{figure}
\begin{center}
\includegraphics[width=15.0cm,clip]{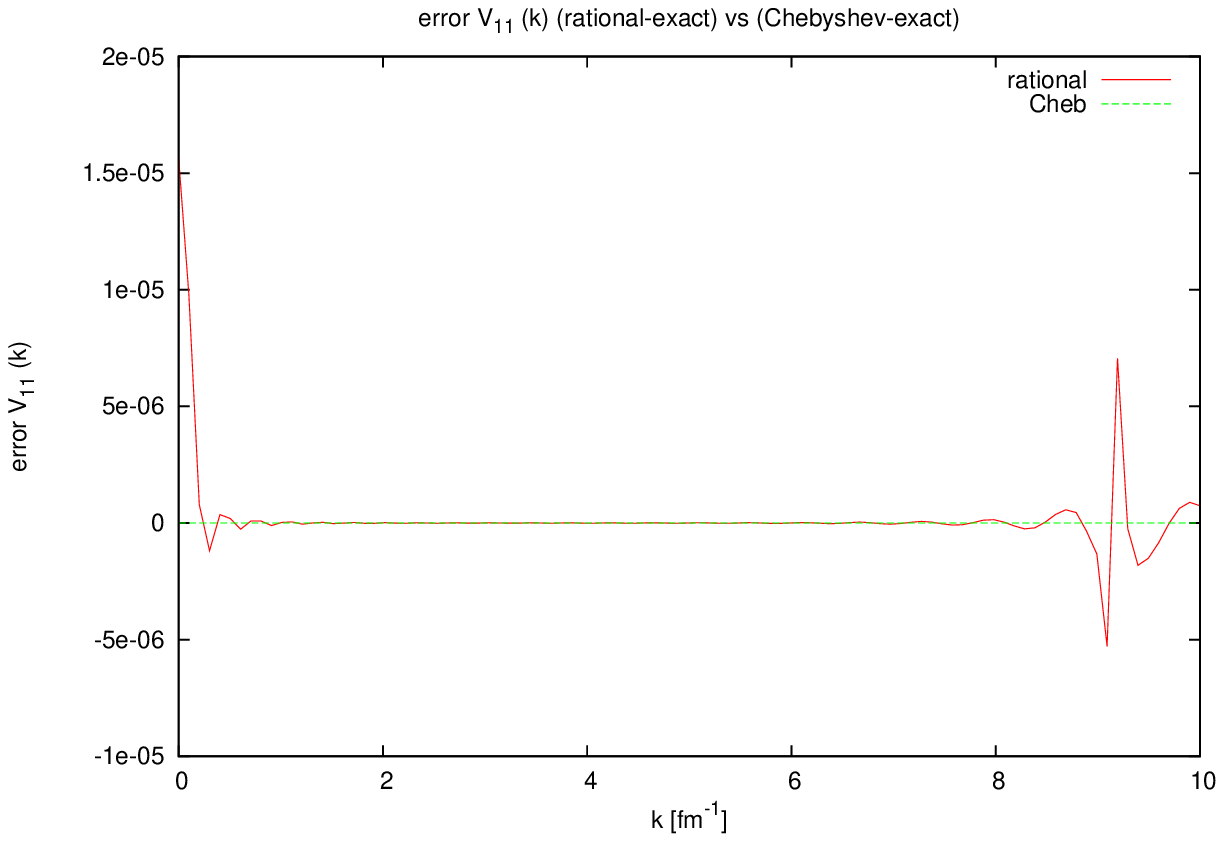}
\caption[Short caption for figure 13]{\label{labelFig13} 
$\Delta V_{11b\, rational}(k),\,
\Delta V_{11b\, Chebyshev}(k)$. 
}
\end{center}
\label{fig.13}
\end{figure}

\begin{figure}
\begin{center}
\includegraphics[width=15.0cm,clip]{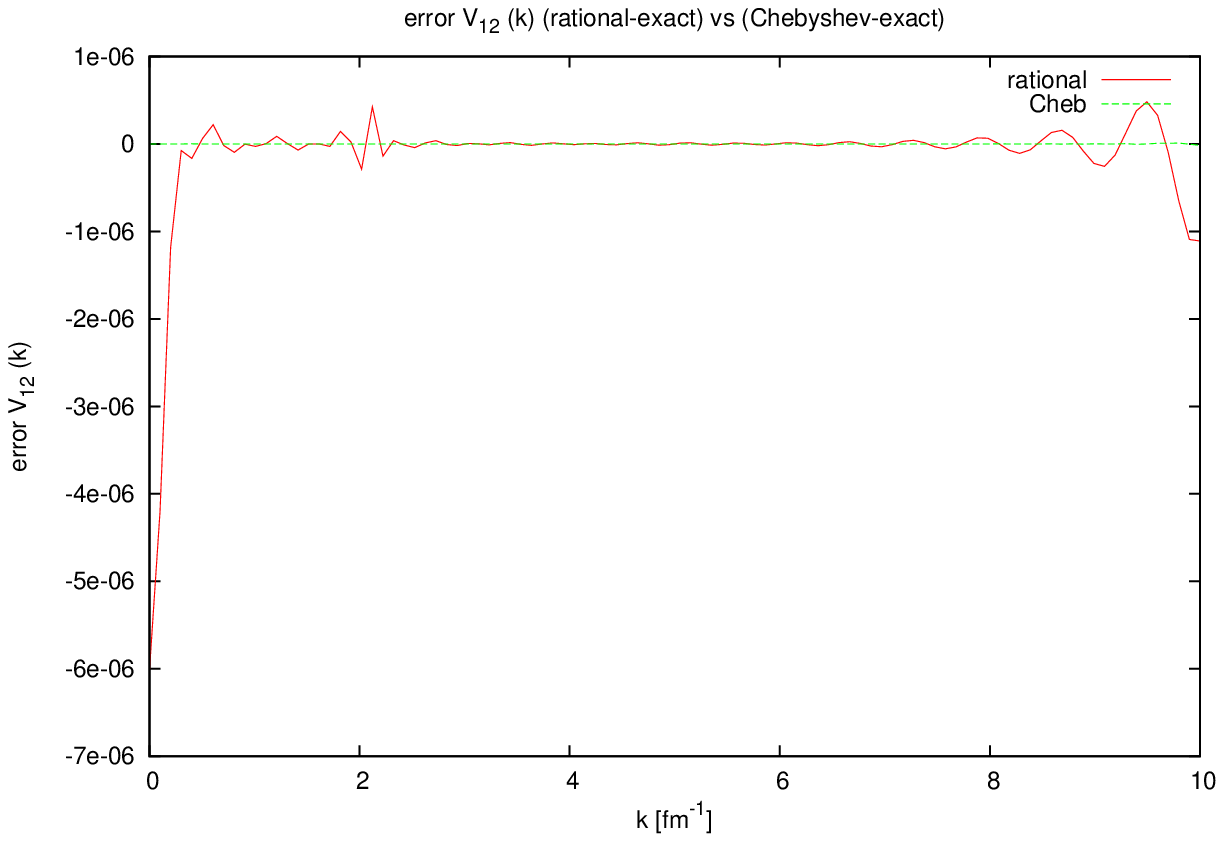}
\caption[Short caption for figure 14]{\label{labelFig14}
$\Delta V_{12b\, rational}(k),\,
\Delta V_{12b\, Chebyshev}(k)$.
}
\end{center}
\label{fig.14}
\end{figure}

\begin{figure}
\begin{center}
\includegraphics[width=15.0cm,clip]{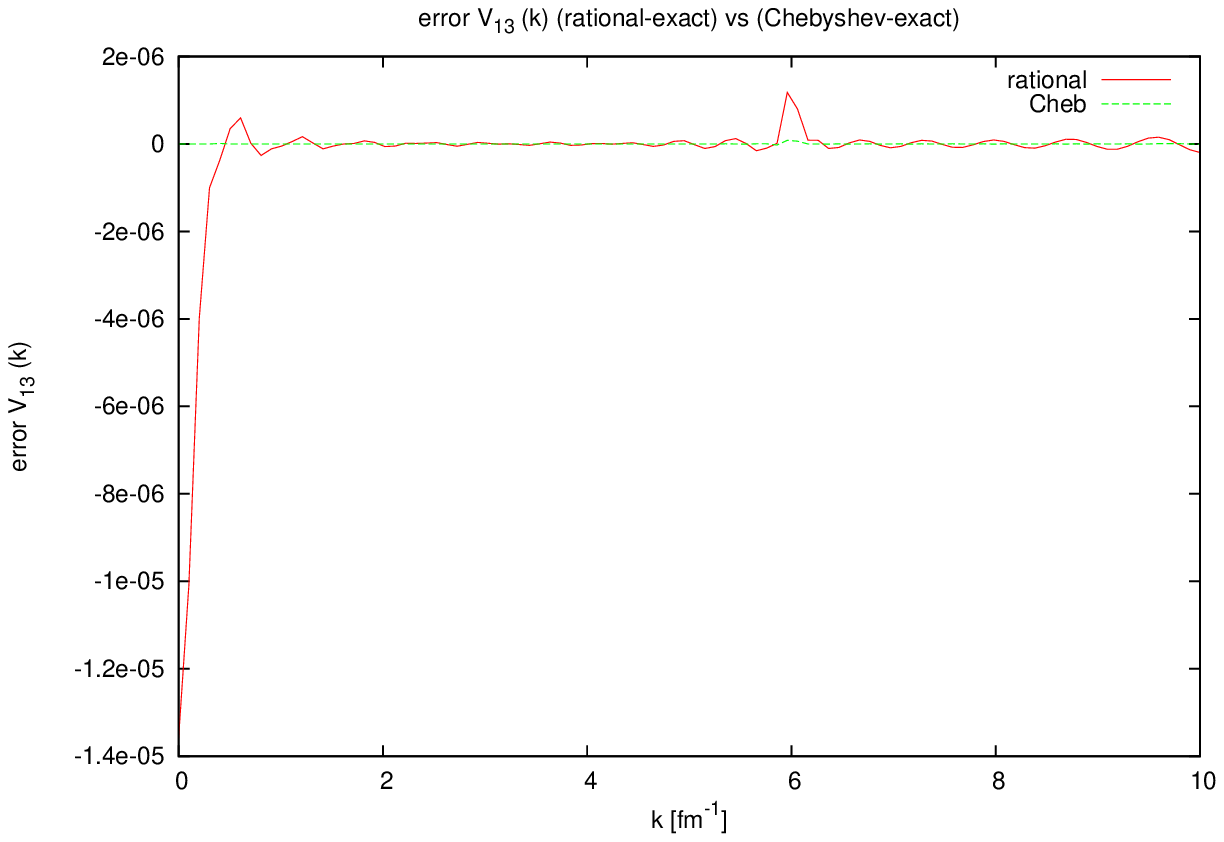}
\caption[Short caption for figure 15]{\label{labelFig15} 
$\Delta V_{13b\, rational}(k),\,
\Delta V_{13b\, Chebyshev}(k)$. 
}
\end{center}
\label{fig.15}
\end{figure}

\begin{figure}
\begin{center}
\includegraphics[width=15.0cm,clip]{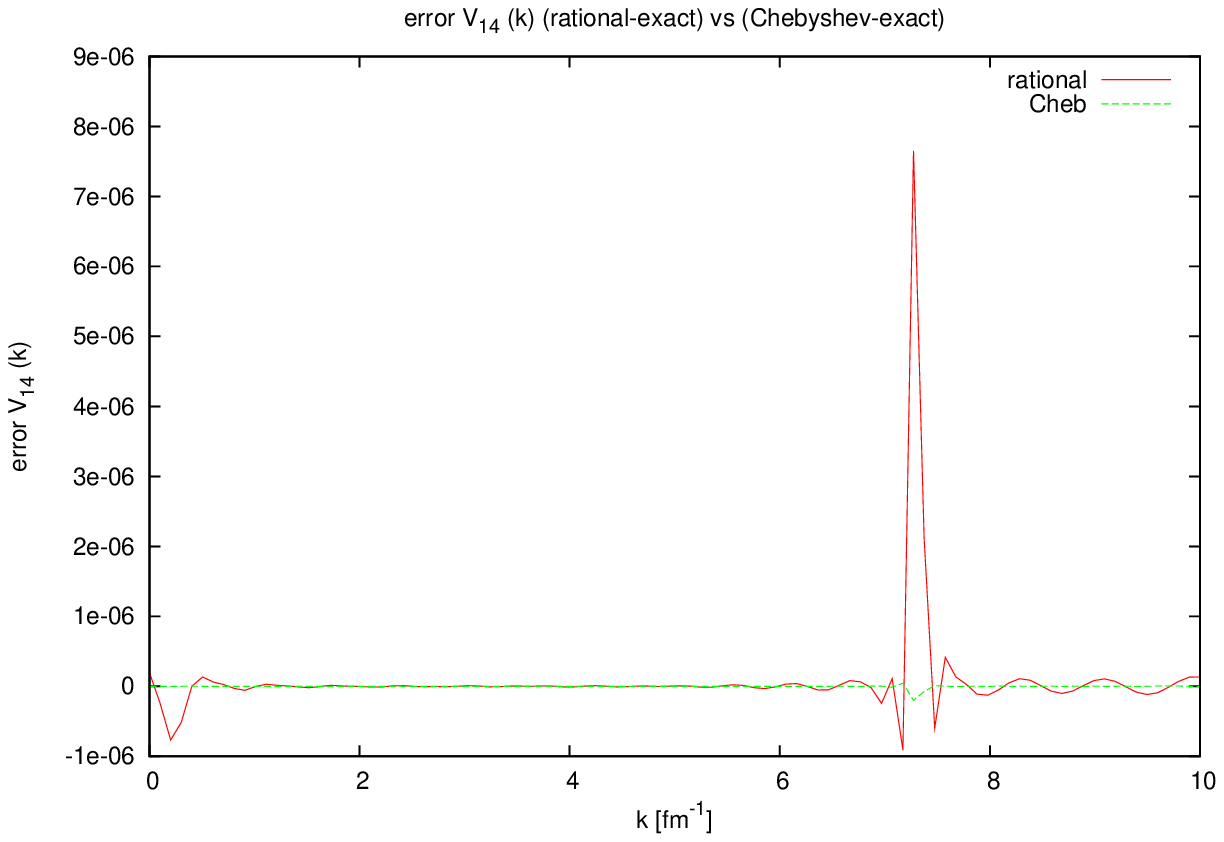}
\caption[Short caption for figure 16]{\label{labelFig16}
$\Delta V_{14b\, rational}(k),\,
\Delta V_{14b\, Chebyshev}(k)$.
}
\end{center}
\label{fig.16}
\end{figure}

\begin{figure}
\begin{center}
\includegraphics[width=15.0cm,clip]{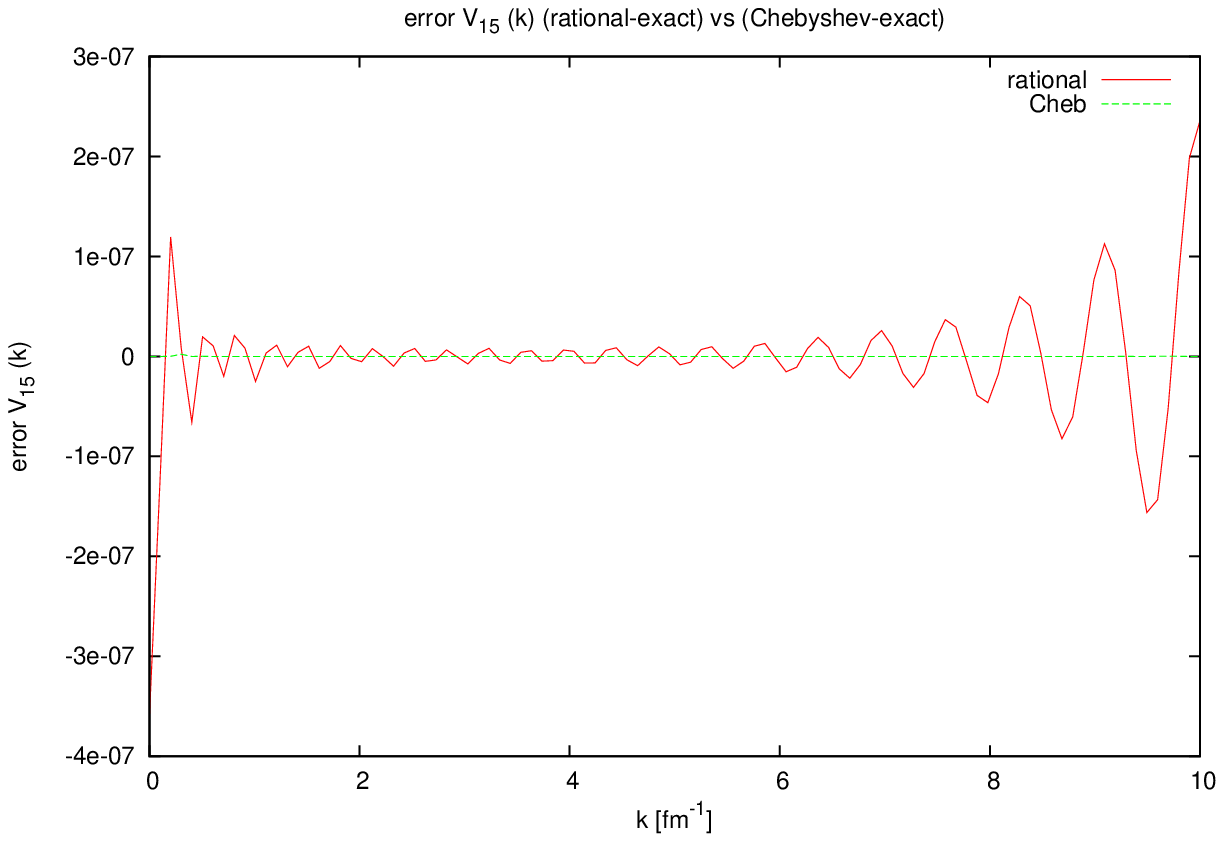}
\caption[Short caption for figure 17]{\label{labelFig17} 
$\Delta V_{15\, rational}(k),\,
\Delta V_{15\, Chebyshev}(k)$. 
}
\end{center}
\label{fig.17}
\end{figure}

\begin{figure}
\begin{center}
\includegraphics[width=15.0cm,clip]{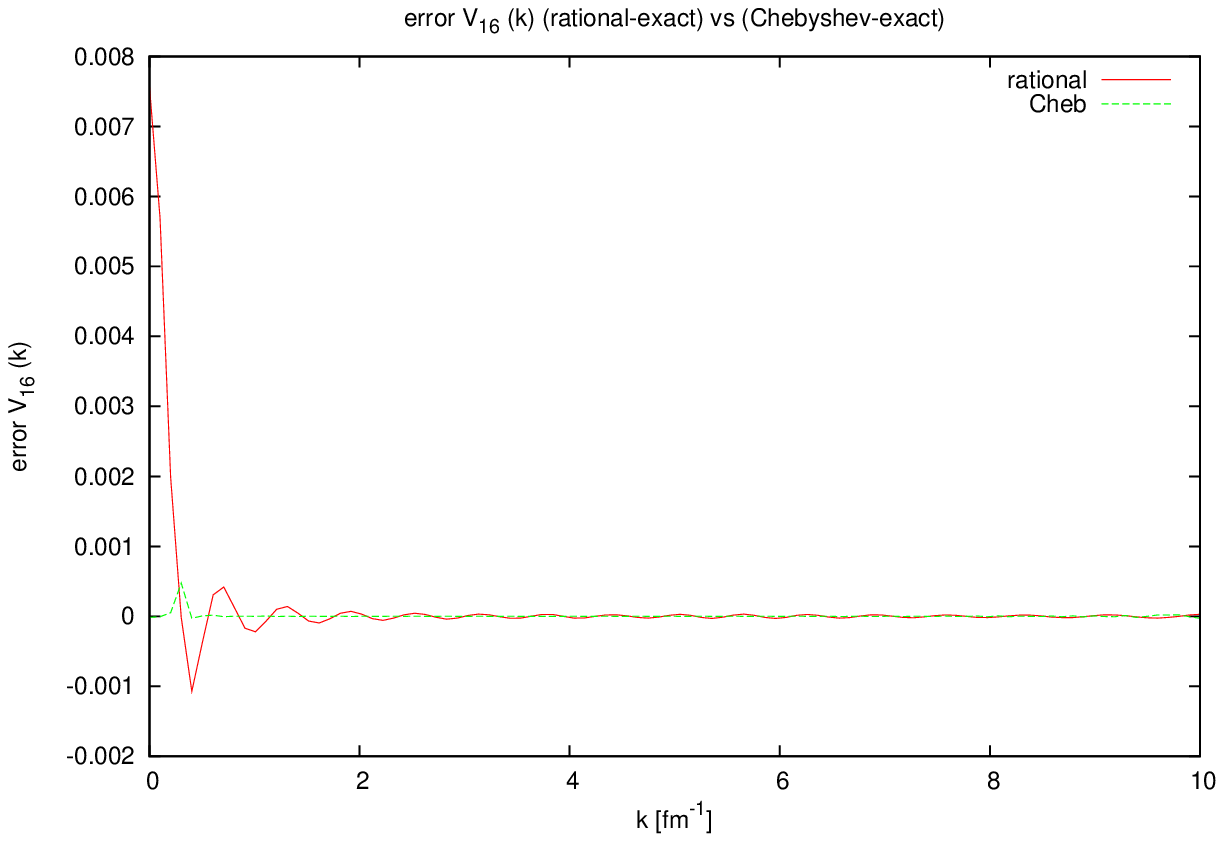}
\caption[Short caption for figure 18]{\label{labelFig18}
$\Delta V_{16\, rational}(k),\,
\Delta V_{16\, Chebyshev}(k)$.
}
\end{center}
\label{fig.18}
\end{figure}

\clearpage

\begin{figure}
\begin{center}
\includegraphics[width=15.0cm,clip]{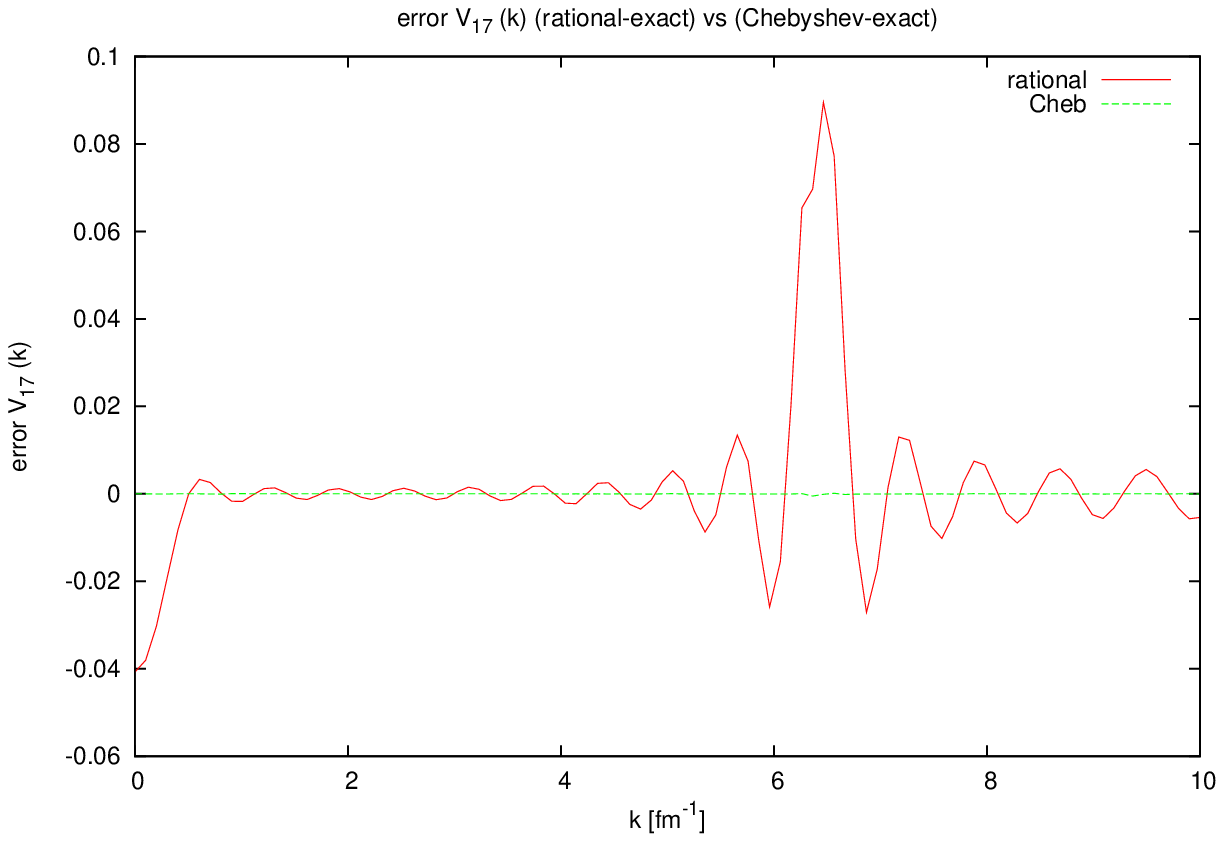}
\caption[Short caption for figure 19]{\label{labelFig19} 
$\Delta V_{17\, rational}(k),\,
\Delta V_{17\, Chebyshev}(k)$. 
}
\end{center}
\label{fig.19}
\end{figure}

\begin{figure}
\begin{center}
\includegraphics[width=15.0cm,clip]{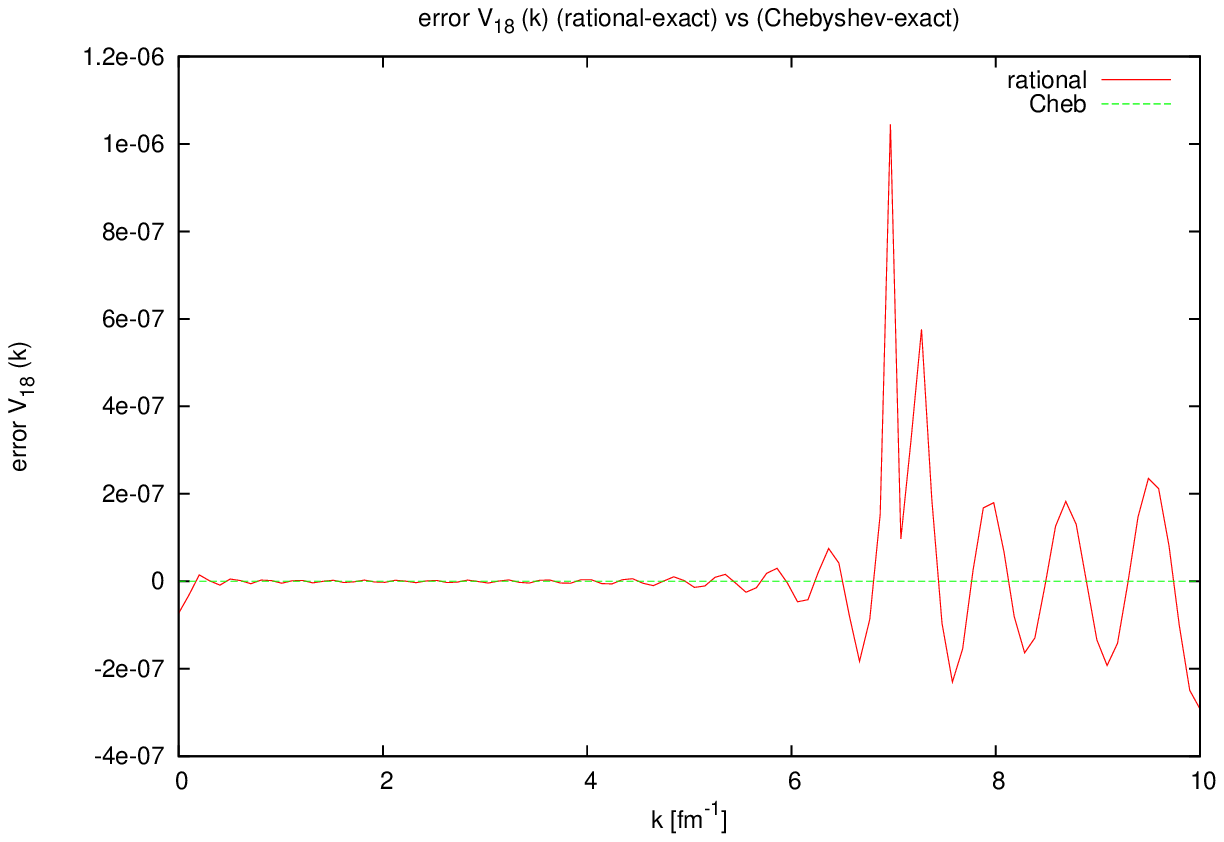}
\caption[Short caption for figure 20]{\label{labelFig20}
$\Delta V_{18\, rational}(k),\, 
\Delta V_{18\, Chebyshev}(k)$.
}
\end{center}
\label{fig.20}
\end{figure}

\begin{figure}
\begin{center}
\includegraphics[width=15.0cm,clip]{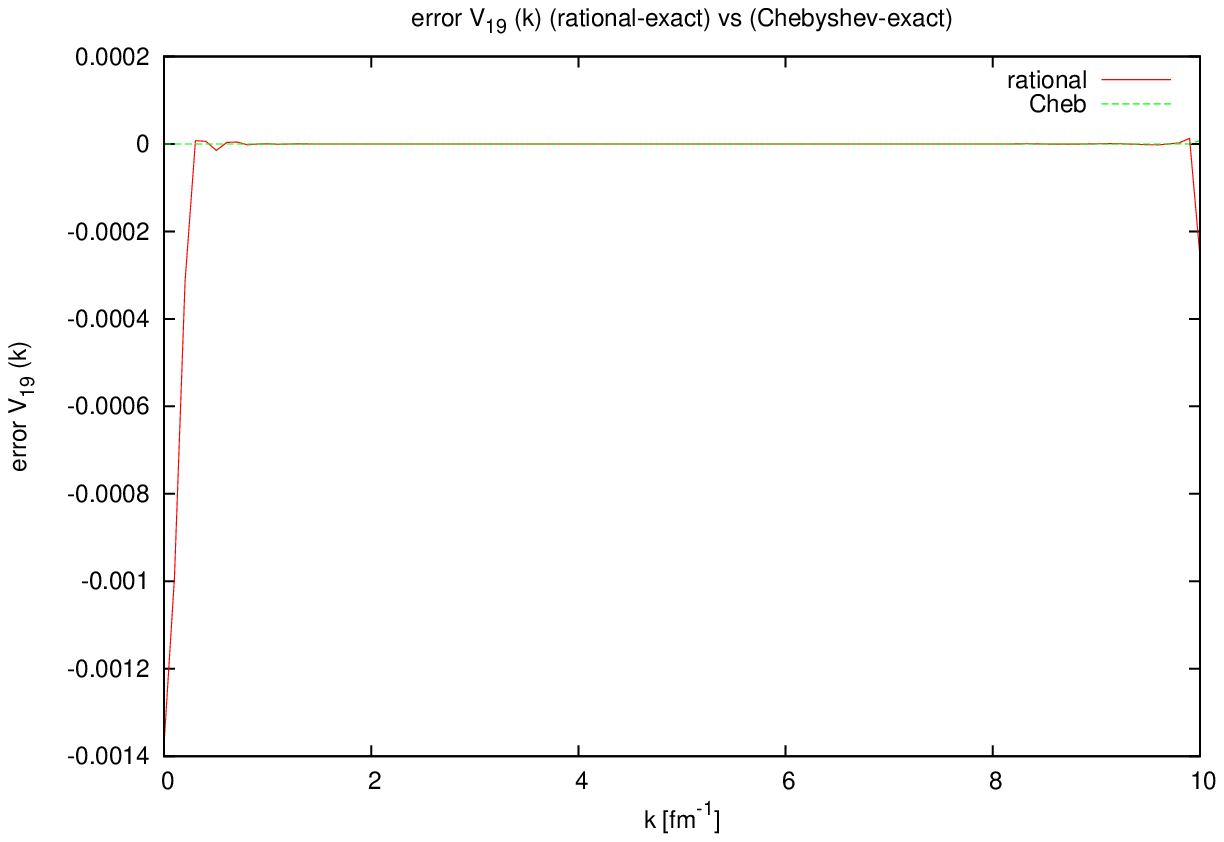}
\caption[Short caption for figure 21]{\label{labelFig21} 
$\Delta V_{9a\, rational}(k),\,
\Delta V_{9a\, Chebyshev}(k)$. 
}
\end{center}
\label{fig.21}
\end{figure}

\begin{figure}
\begin{center}
\includegraphics[width=15.0cm,clip]{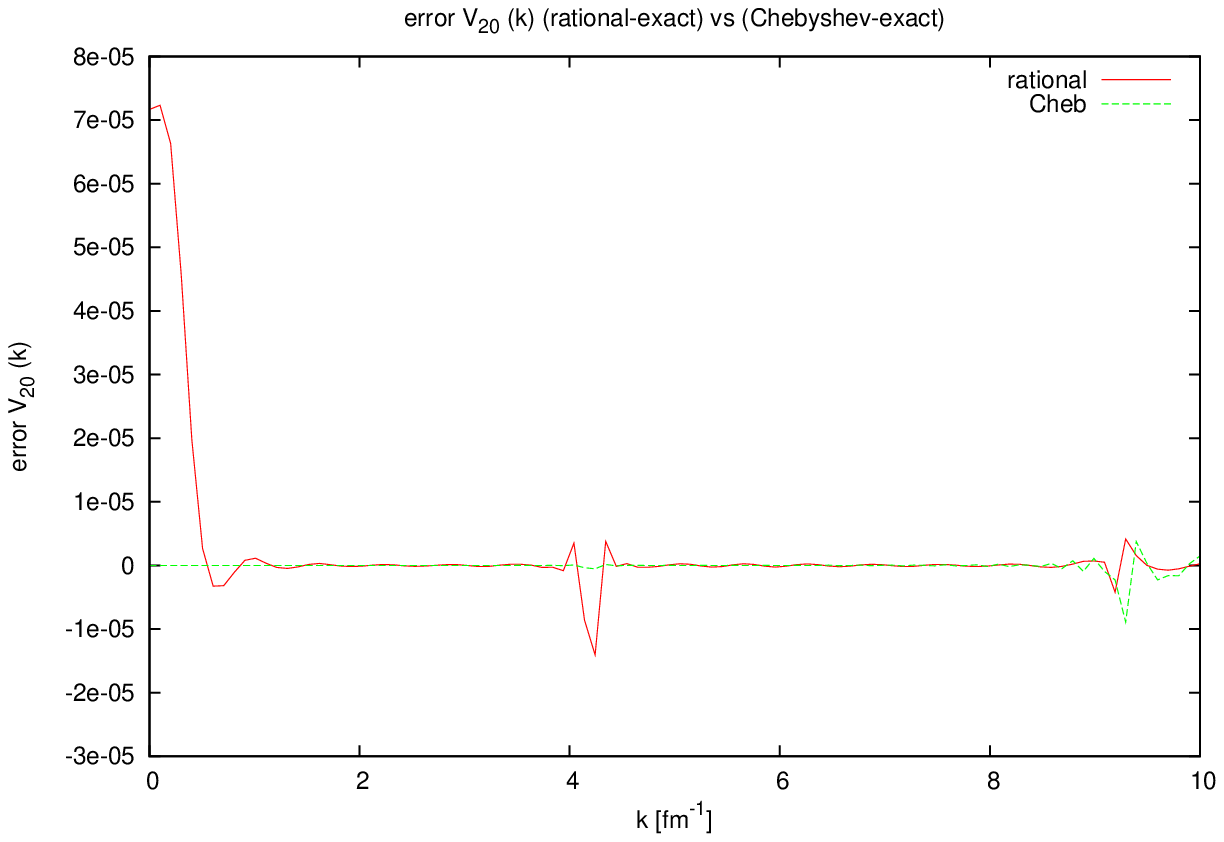}
\caption[Short caption for figure 22]{\label{labelFig22}
$\Delta V_{10a\, rational}(k),\,
\Delta V_{10a\, Chebyshev}(k)$.
}
\end{center}
\label{fig.22}
\end{figure}

\begin{figure}
\begin{center}
\includegraphics[width=15.0cm,clip]{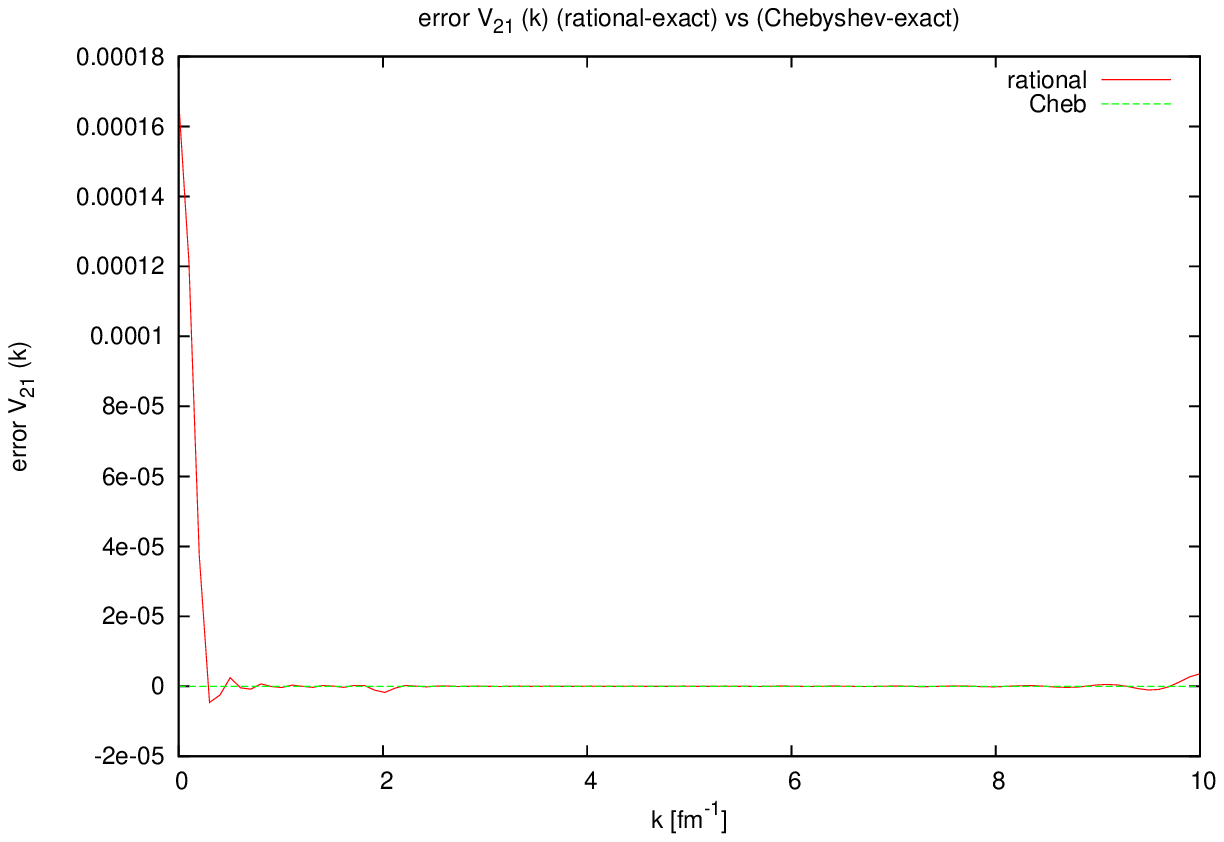}
\caption[Short caption for figure 23]{\label{labelFig23} 
$\Delta V_{11a\, rational}(k),\,
\Delta V_{11a\, Chebyshev}(k)$. 
}
\end{center}
\label{fig.23}
\end{figure}

\begin{figure}
\begin{center}
\includegraphics[width=15.0cm,clip]{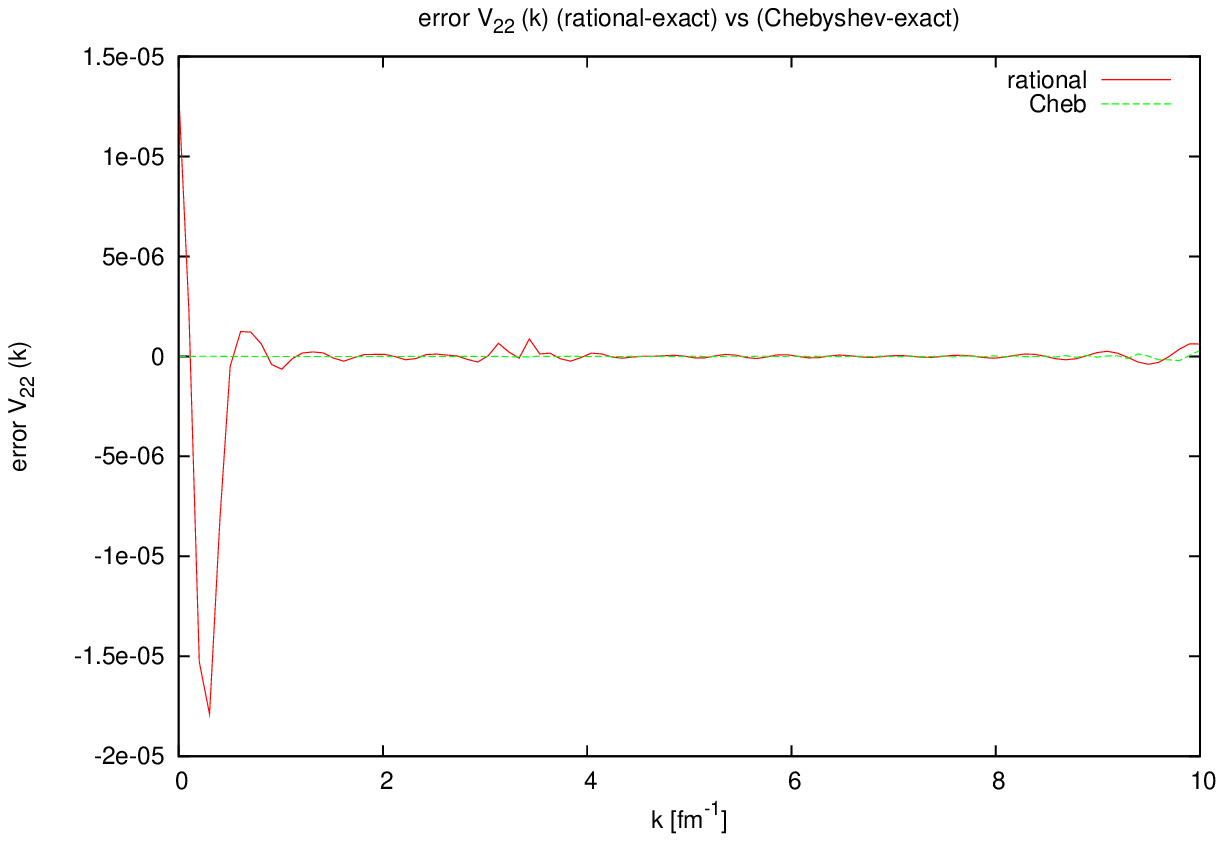}
\caption[Short caption for figure 24]{\label{labelFig24}
$\Delta V_{12a\, rational}(k),\,
\Delta V_{12a\, Chebyshev}(k)$.
}
\end{center}
\label{fig.24}
\end{figure}

\begin{figure}
\begin{center}
\includegraphics[width=15.0cm,clip]{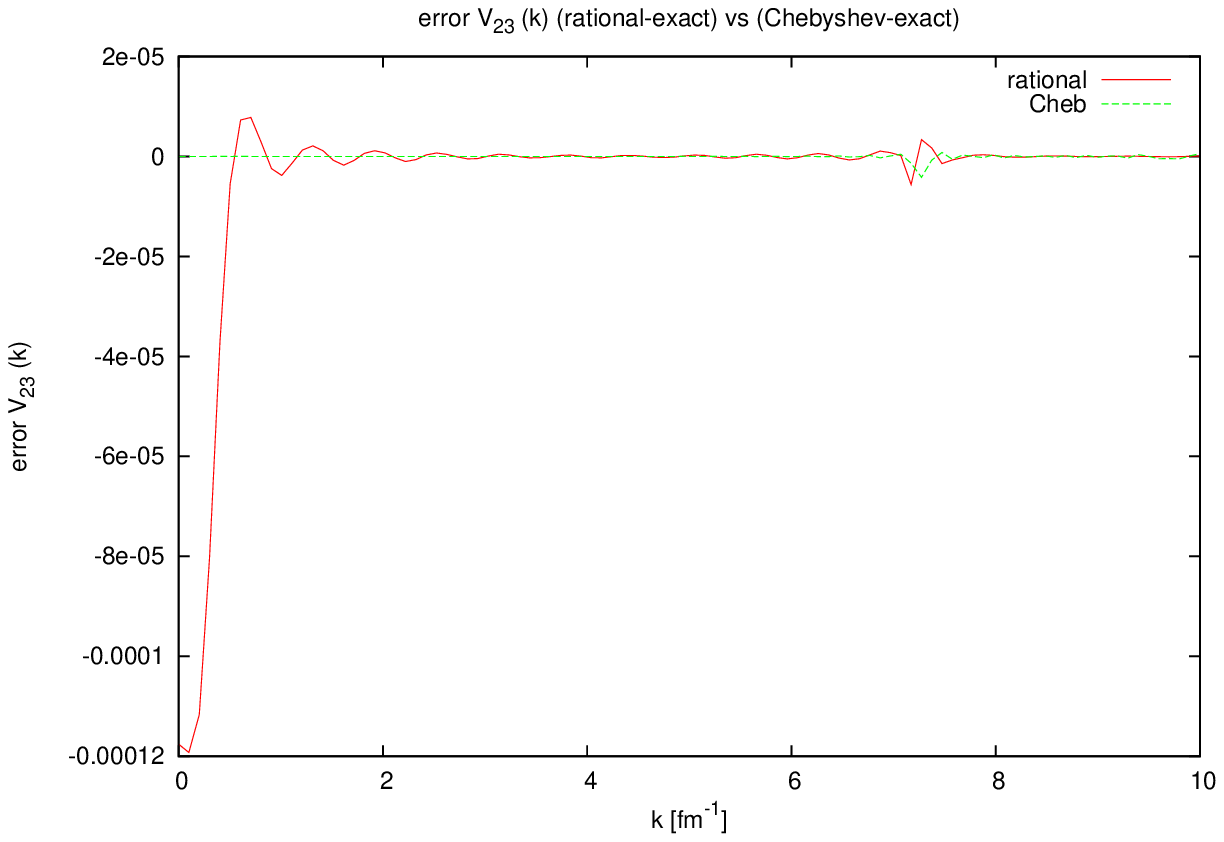}
\caption[Short caption for figure 25]{\label{labelFig25} 
$\Delta V_{13a\, rational}(k),\,
\Delta V_{13a\, Chebyshev}(k)$. 
}
\end{center}
\label{fig.25}
\end{figure}

\begin{figure}
\begin{center}
\includegraphics[width=15.0cm,clip]{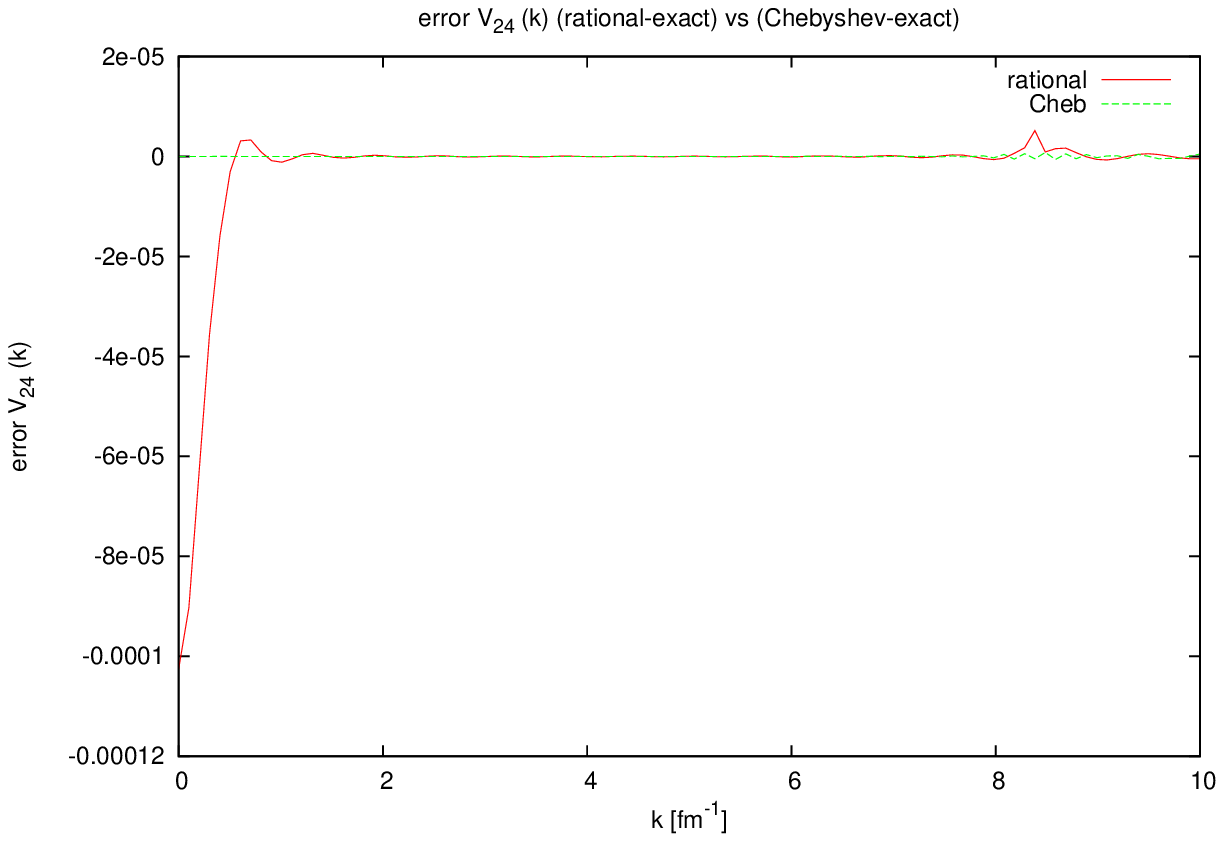}
\caption[Short caption for figure 26]{\label{labelFig26}
$\Delta V_{14a\, rational}(k),\,
\Delta V_{14a\, Chebyshev}(k)$.
}
\end{center}
\label{fig.26}
\end{figure}

\vfill\eject


\begin{thebibliography}{00}

\bibitem{v18} R.B. Wiringa, V.G.J. Stoks, R. Schiavilla, 
Phys. Rev. {\bf C}51(1995)38.
\bibitem{reid} V.G.J. Stoks, R.A.M. Klomp, C.P.F. Terheggen, 
J. J.  de Swart, Phys. Rev. {\bf C}49(1994)2950.
\bibitem{cdbonn}
R. Machleidt, Phys. Rev. {\bf C}63(2001)024001.
\bibitem{huber} W. Gl\"ockle,  H. Witala, D. H\"uber, H. Kamada, J. Golak, Phys. Rep. {\bf 274}\
, 107 (1986).
\bibitem{thesis2011} S. Veerasamy, University of Iowa Thesis, 2011.
\bibitem{broucke} R. Broucke, Communications of the Association for 
Computing Machinery, 16,(1973)254.
\bibitem{keister} B. D. Keister and W. N. Polyzou,
J. Comp. Phys. 134(1997),231. 
\bibitem{charlotte} J. Golak,  
W. Gl\"ockle , R. Skibi\'nski, H. Wita{\l}a, D. Rozpkedzik, K. Topolnicki, 
I. Fachruddin, 
Ch. Elster, A. Nogga, Phys. Rev. {\bf C}81(2010)034006. 
\bibitem{wiringa} Private communication with Robert Wiringa.
\bibitem{ce11} http://arxiv.org/abs/arXiv:0911.4173 .
\bibitem{gradshteyn1} I. S. Gradshteyn and I. M. Ryzhik,
{\it Tables of Integrals, Series, and Products}, Associated Press,
(1965), page 711 eq. 6.621.
\bibitem{ak}N. Auerbach, J. H\"ufner, A. K. Kerman, and C. M. Shakin,  \
Reviews of Modern Physics, {\bf 44},(1972)48. 
\bibitem{gradshteyn2} I. S. Gradshteyn and I. M. Ryzhik,
{\it Tables of Integrals, Series, and Products}, Associated Press,
(1965), page 583 eq. 4.381,page 605 eq. 4.401.

\end{thebibliography}
\end{document}